\documentclass[12pt]{article}
\usepackage{amsmath}
\usepackage{amssymb,amscd}
\usepackage{bm}
\usepackage{wrapfig}
\usepackage{color}
\usepackage{theorem}

\newtheorem{theorem}{Theorem}[section]
 \newtheorem{definition}[theorem]{Definition}
 \newtheorem{proposition}[theorem]{Proposition}

{\theorembodyfont{\normalfont}
 \newtheorem{example}{Example}[section]
 \newtheorem{remark}{Remark}[section]
 }


\newlength{\minitwocolumn}
\catcode`\@=11

\def\ps@myheadings{\let\@mkboth\@gobbletwo
\def\@oddhead{\hbox{}
\rightmark\hfil\eightrm\thepage}
\def\@oddfoot{}\def\@evenhead{\eightrm\thepage\hfil
\leftmark\hbox{}}\def\@evenfoot{}
\def\sectionmark##1{}\def\subsectionmark##1{}}

\renewcommand{\theequation}{\thesection.\arabic{equation}}

\newenvironment{proof}[1][Proof]{\begin{trivlist}
\item[\hskip \labelsep {\bfseries #1}]}{\end{trivlist}}
\newcommand{\qed}{\nobreak \ifvmode \relax \else
      \ifdim\lastskip<1.5em \hskip-\lastskip
      \hskip1.5em plus0em minus0.5em \fi \nobreak
      \vrule height0.75em width0.5em depth0.25em\fi}

\newcommand\hQ{\mbox{\boldmath $Q$}}

\newcommand{\rd}{\overleftarrow{\partial}}
\newcommand{\ld}{\overrightarrow{\partial}}
\newcommand{\sbv}[2]{{\{{{#1},{#2}}\}}}

\newcommand{\mbv}[2]{{\{{{#1},{#2}}\}_{BV}}}

\newcommand{\bracket}[2]{\langle #1\,,#2\rangle}

\def\bx{\mbox{\boldmath $x$}}
\def\bxi{\mbox{\boldmath $\xi$}}
\def\bq{\mbox{\boldmath $q$}}
\def\bp{\mbox{\boldmath $p$}}

\def\bbx{\mbox{\boldmath $x$}}
\def\bbxi{\mbox{\boldmath $\xi$}}

\def\bbq{\mbox{\boldmath $q$}}
\def\bbp{\mbox{\boldmath $p$}}

\def\bbd{\mbox{\boldmath $d$}}

\def\bbz{\mbox{\boldmath $z$}}

\def\bomega{\mbox{\boldmath $\omega$}}

\def\bR{\mbox{\boldmath $R$}}

\def\bZ{\mbox{\boldmath $Z$}}

\def\bPhi{\mbox{\boldmath $\Phi$}}

\newcommand{\calL}{{\cal L}}
\newcommand{\calM}{{\cal M}}
\newcommand{\calN}{{\cal N}}

\newcommand{\calX}{{\cal X}}

\newcommand{\Map}{{\rm Map}}
\newcommand{\ev}{{\rm ev}}

\newcommand{\nom}{\nonumber}
\newcommand{\beq}{\begin{equation}}
\newcommand{\eeq}{\end{equation}}
\newcommand{\bea}{\begin{eqnarray*}}
\newcommand{\eea}{\end{eqnarray*}}
\newcommand{\beqa}{\begin{eqnarray}}
\newcommand{\eeqa}{\end{eqnarray}}

\newcommand{\bJ}{\boldsymbol{J}}

\newcommand{\del}{\partial}

\newcommand{\courant}[2]{{[{{#1},{#2}}]_D}}
\newcommand{\couranth}[2]{{[{{#1},{#2}}]_H}}
\newcommand{\courantr}[2]{{[{{#1},{#2}}]_R^{\pi}}}

\newcommand{\be}{\boldsymbol{e}}

\newcommand\proj{{pr}}
\newcommand{\tilpi}{{\tilde{\proj}}}
\newcommand{\hatpi}{{\widehat{\proj}}}
\newcommand\hatcalL{{\widehat{\calL}}}

\newcommand{\tx}{\tilde{x}}
\newcommand{\ty}{\tilde{y}}

\newcommand{\teta}{\tilde{\eta}}
\newcommand{\tpar}{\tilde{\partial}}

\def\bqpr{\mbox{\boldmath $q^{\prime}$}}
\def\bppr{\mbox{\boldmath $p^{\prime}$}}
\def\bbqpr{\mbox{\boldmath $q^{\prime}$}}
\def\bbppr{\mbox{\boldmath $p^{\prime}$}}
\def\bxipr{\mbox{\boldmath $\xi^{\prime}$}}
\def\bxpr{\mbox{\boldmath $x^{\prime}$}}
\def\bbxpr{\mbox{\boldmath $x^{\prime}$}}

\newcommand\eeqref{\eqref}
\newcommand\qone{{\eta}}

\begin{document}


\baselineskip 0.7cm



\begin{titlepage}
\begin{flushright}
\null \hfill Preprint TU-1003\\[3em]
\end{flushright}

\begin{center}
{\Large \bf
Topological Membranes, Current Algebras 
and\\ H-flux - R-flux Duality 
based on Courant Algebroids
}
\vskip 1.2cm
Taiki Bessho${}^{a,}$\footnote{E-mail:\
tbessho@tuhep.phys.tohoku.ac.jp}, Marc Andre Heller${}^{a,}$\footnote{E-mail:\
heller@tuhep.phys.tohoku.ac.jp}, Noriaki Ikeda${}^{b,}$\footnote{E-mail:\
nikeda@se.ritsumei.ac.jp
}
 ~and Satoshi Watamura${}^{a,}$\footnote{E-mail:\ watamura@tuhep.phys.tohoku.ac.jp}  %

\vskip 0.4cm
{

\it
${}^a$
Particle Theory and Cosmology Group, \\
Department of Physics, Graduate School of Science, \\
Tohoku University \\
Aoba-ku, Sendai 980-8578, Japan \\ 

\vskip 0.4cm
${}^b$
Department of Mathematical Sciences,
Ritsumeikan University \\
Kusatsu, Shiga 525-8577, Japan \\

}
\vskip 0.4cm


\vskip 1.5cm

\begin{abstract}
We construct a topological sigma model and a current algebra based on a Courant algebroid structure on a Poisson manifold. In order to construct models, we reformulate the Poisson Courant algebroid by supergeometric construction on a QP-manifold. A new duality of Courant algebroids which transforms $H$-flux and $R$-flux is proposed, where the transformation is interpreted as a canonical transformation of a graded symplectic manifold.
\end{abstract}
\end{center}
\end{titlepage}

\setcounter{page}{2}


\rm

\section{Introduction}
\noindent
There exist various dualities in string theory. Among them, T-duality is 
directly connected with the geometry of the target space and 
thus has to be a characteristic property of stringy geometry. 


One of the formulations to analyze T-duality 
is the approach of 
{doubled geometry, 
which 
has 
manifest $O(d,d)$ invariance},
and there, the existence of 
so-called nongeometric fluxes has been proposed \cite{Hull:2004in}.
On the other hand, the fluxes $H$, $F$, $Q$ and $R$
and their transformations 
have also been conjectured from T-duality analysis in supergravity compactification scenario
\cite{Shelton:2005cf,Grana:2008yw}.
It has been proposed that 
T-duality converts $H$-, $F$-, $Q$- and $R$-fluxes into each other.
Recently, there are further developments related to T-duality. 
{Double field theory \cite{Hull:2009mi} is a manifestly $O(d,d)$ covariant
field theory which allows also 
for T-duality along non-isometry directions.}
Examples for other developments are
the branes as sources for $Q$- and $R$-fluxes 
\cite{deBoer:2012ma, Hassler:2013wsa}
and the $\beta$-supergravity \cite{Andriot:2013xca}.
The topological T-duality 
\cite{Bouwknegt:2003vb,Bouwknegt:2004tr}
is also proposed to analyze T-duality with flux.
However, the background geometric structures for nongeometric fluxes 
are not well understood.


A background geometry in string theory with NS $H$-flux 
\cite{Bouwknegt:2010zz} is known to be a \textit{Courant algebroid} 
\cite{Courant,lwx}, and the standard Courant algebroid of 
the generalized tangent bundle $TM\oplus T^*M$ is of particular interest in the framework of generalized geometry \cite{Hitchin:2004ut,Gualtieri:2003dx}.
The T-duality on the $H$-flux is well understood 
as an automorphism on the standard Courant algebroid 
if $\iota_X \iota_Y H = 0$ \cite{Cavalcanti:2011wu}.
However, 
we cannot simultaneously introduce all degrees of freedom 
of $H$-, $F$-, $Q$-, $R$- fluxes 
as deformation of the Courant algebroid.
The only independent deformation in the exact Courant 
algebroid is a $3$-form ($H$-flux) 
degree of freedom \cite{Severa}.

Recently, the Courant algebroid on a Poisson manifold, 
i.e. the Poisson Courant algebroid, has been 
introduced in \cite{Asakawa:2014kua} as a geometric object
for a background with $R$-flux.
It is shown that the nontrivial flux $R$ of a $3$-vector can be introduced consistently on a Poisson manifold 
as a deformation of
the Courant algebroid.
It is the 'contravariant object' \cite{Fernandes00} with respect to  
the standard Courant algebroid, which is the exchange of 
$T^{*}M$ with $TM$ and $H$-flux with $R$-flux.
The T-duality on the $R$-flux has also been analyzed and 
it has been shown that the duality of $R$-flux with $Q$-flux is also 
understood as an automorphism on the Poisson Courant algebroid \cite{Asakawa:2015aia}.

In this paper, we analyze the geometric structure of
the Poisson Courant algebroid and a duality between $H$-flux and 
$R$-flux, which we call flux duality, in detail. 
We also construct the corresponding worldvolume theories,  
a topological sigma model and a current algebra with the structure 
of this Poisson Courant algebroid.

We first discuss the mathematical features and 
some structural correspondences between the standard Courant algebroid 
with $H$-flux and 
the Poisson Courant algebroid with $R$-flux.
By analyzing both coboundary operators, we generalize
the duality between the de Rham cohomology and 
the Poisson cohomology as the background algebraic structure.
Moreover, we have a new interpretation of this duality as a canonical transformation
on a graded symplectic manifold\footnote{
{T-duality has been formulated as a canonical transformation
on the string phase space in \cite{Alvarez:1994wj, Klimcik:1994gi}.
Canonical transformations in this paper are defined on 
a graded target manifold.}},
and 
we formulate a flux duality,
a duality between the $H$-flux and $R$-flux.

Then, we discuss field theoretic realizations of the Poisson Courant algebroid as 
a symmetry; we construct a topological sigma model and a current algebra. 
To this end, first we reformulate the Courant algebroid in terms of supergeometry. 
The construction of the Courant algebroid by using supergeometry and 
the derived brackets are introduced in \cite{Roytenberg99}. 
This formulation uses a so-called QP-manifold, 
a differential graded symplectic manifold
\cite{Schwarz:1992nx,Schwarz:1992gs}.
The advantage of the use of supergeometry is 
that 
the topological sigma model and the current algebra are 
constructed straightforwardly
 from the supergeometric data.
The general theories are known as 
the AKSZ construction of topological sigma models \cite{Alexandrov:1995kv}
and the supergeometric BFV formulation of current algebras 
\cite{Ikeda:2013vga}.

It is known that the AKSZ sigma model in three dimensions generally has the structure of 
a Courant algebroid \cite{Ikeda:2002wh,Hofman:2002rv,Roytenberg:2006qz}. 
Physically, this is a theory of a topological membrane.
Following general arguments, we construct  
a topological sigma model from the Poisson Courant 
algebroid in three dimensions. 
When the three-dimensional world volume has a boundary,
i.e. when we consider the open membrane,
we obtain a two-dimensional boundary sigma model \`{a} la WZW.
This is the Poisson sigma model with $R$-flux
on the Dirac structure of a Poisson Courant algebroid.
From the point of view of the sigma model, 
T-duality is changing the boundary conditions 
of the topological membranes.
There is an approach with a similar concept proposed in 
\cite{Mylonas:2012pg}. The difference is that our formalism is based on the Poisson 
Courant algebroid.

We also construct a current algebra of the Poisson Courant algebroid on 
loop space, coming from the canonical formulation of the theories on 
($1+1$)-dimensional spacetime $S^1 \times \bR$. 
In the H-flux case, this is
the Alekseev-Strobl current algebra
\cite{Alekseev:2004np}, which has the structure of
the standard Courant algebroid with H-flux as underlying symmetry. 
This type of current algebra can also be reformulated by using the supergeometric construction 
\cite{Ikeda:2011ax,Ikeda:2013vga}. 
Following these general formulations, we construct a corresponding 
current algebra with R-flux.

This paper is organized as follows. 
In section 2, we review the supergeometric construction of Courant algebroids and,
in section 3, we apply it to the Poisson Courant algebroid.
In section 4, we discuss the mathematical structure of the duality of 
the standard Courant algebroid and the Poisson Courant algebroid. 
In section 5, we discuss the meaning of $R$-flux of
the Poisson Courant algebroid from the perspective of double field theory.
In section 6, we review the AKSZ sigma model. Then, we construct a topological 
sigma model of the Poisson Courant algebroid and analyze 
its boundary theories. 
In section 7, we construct the current algebra of the Poisson Courant 
algebroid.
Finally, section 8 is devoted to conclusion and discussion.
In the Appendix, our notation of supergeometry in this paper is summarized.

\section{Courant algebroids and supergeometry}\label{CAandsupergeometry}
\noindent
In this section, we briefly review supergeometry, its definition and related 
terms which are necessary to construct the topological sigma models from the 
standard Courant algebroids using the AKSZ construction in 
section 5. 
Here, we review 
definitions of Courant algebroids in the first subsection. 
Courant algebroids provide the background geometry of T-duality. The second subsection then reviews a differential 
graded symplectic manifold, which is called a QP-manifold. 
In the AKSZ formulation, a QP-manifold is used to construct a
topological sigma model. 
Finally, in the third subsection, 
the supergeometric construction of the Courant algebroids 
from QP-manifolds of degree $2$ is explained.
The formulation is based on the fundamental theorem 
that general Courant algebroids are 
\textit{equivalent} to QP-manifolds of degree $2$.
This short review of the techniques 
involved provides the foundation to flow
into the definition of Poisson Courant algebroids and 
their realization through supergeometric 
construction.
%

\subsection{Courant algebroids}
\noindent
Let us start with recalling the definition of the Courant algebroid.
\begin{definition}
\cite{lwx,Kosmann-Schwarzbach05}
\label{CourantKS}
The \textsl{Courant algebroid} is a vector bundle $E$ over $M$ with 
three operations, a pseudo-Euclidean metric $\bracket{-}{-}$ on 
the fiber, a bundle map $\rho:E \longrightarrow TM$ (called the anchor map),
and a binary bracket $[-,-]_{D}$ (the Dorfman bracket) on the space 
of sections $\Gamma (E)$,
which satisfy the following conditions:
\begin{eqnarray}\label{defcou1}
\label{cou1}
1)&[e_1 ,[e_2 , e_3]_{D}]_{D}
&=~[[e_1 ,e_2]_{D},  e_3]_{D}+[e_2 ,[e_1, e_3]_{D}]_{D},
\nonumber
\\
\label{cou2}2)& \rho(e_1)\bracket{e_2}{e_3}&=~
\bracket{[e_1,e_2]_{D}}{e_3}+\bracket{e_2}{[e_1,e_3]_{D}},
\nonumber
\\
\label{cou3}3)& \rho(e_1)\bracket{e_2}{e_3}
&=~\bracket{e_1}{[e_2,e_3]_{D}+[e_3,e_2]_{D}},
\nonumber
\end{eqnarray}
where $e_1 ,e_2 , e_3 \in \Gamma (E)$.
\end{definition}
Regarding its application to string theory, Courant algebroids
appeared in the context of generalized geometry 
\cite{Hitchin:2004ut,Gualtieri:2003dx}. 
In this case, the vector bundle is the direct sum of tangent bundle 
and cotangent bundle $E=TM\oplus T^*M$
and is introduced as an extension of the
Lie algebroid of tangent vectors. 
We call such a Courant algebroid 
the \textit{standard Courant algebroid}:
\begin{definition}
\label{standardCA}
The \textsl{standard Courant algebroid} is a Courant algebroid 
as defined above, where we take $E = TM \oplus T^*M$. 
The anchor is the natural projection $\rho:TM\oplus T^*M \longrightarrow TM$. 
The operations of the Courant algebroid
are as follows:
\begin{align}
	\bracket{X + \alpha}{Y + \beta} &= \iota_X \beta + \iota_Y \alpha, \notag \\
\rho(X+\alpha) &= X, \notag \\
\courant{X + \alpha}{Y + \beta} &= [X, Y] + L_X \beta - \iota_Y d \alpha,
\notag
\end{align}
for sections $X + \alpha, Y + \beta \in \Gamma(TM \oplus T^*M)$, 
where $X,Y$ are vector fields and $\alpha,\beta$ are $1$-forms. 
\end{definition}

In string theory, there exists a $3$-form flux, 
which is usually called $H$-flux.
In the case of a compactification, this flux can be nonvanishing in general.
We can make a deformation of the standard Courant algebroid
by a closed $3$-form $H$, which preserves the Courant algebroid conditions.
We call such a deformed Courant algebroid a Courant algebroid with H-flux.
\begin{definition}
The \textsl{standard Courant algebroid with H-flux} is the Courant 
algebroid which contains the same inner product $\bracket{-}{-}$ and
anchor map $\rho:TM  \oplus T^*M \longrightarrow TM$ as 
the standard Courant algebroid. 
The Dorfman bracket is deformed by the 3-form flux to the bracket defined by
\begin{eqnarray}
[{X + \alpha},{Y + \beta}]_H = [X, Y] + L_X \beta - \iota_Y d \alpha
+ \iota_X \iota_Y H. 
\label{standardDorfmanbracket}
\end{eqnarray} 
\end{definition}
The Courant algebroids on $TM \oplus T^*M$,
more precisely, exact Courant algebroids are classified by 
$H^3_{}(M, \bR)$ \cite{Severa}.
This means that we can only introduce the $H$-flux deformation
as independent degree of freedom 
among all fluxes in the standard Courant algebroid.

\subsection{Supergeometric construction}
\noindent
In the following, we review the supergeometric formulation of 
the Courant algebroids based on a so-called QP-manifold. 
Here, the structures of a Courant algebroid 
in the previous section is reconstructed by
the supergeometric method.

First, we give a definition of a graded manifold.
A graded manifold $\calM$ is a ringed space,
whose structure sheaf is a
\bZ-graded commutative
algebra over an ordinary smooth manifold $M$.
The grading is compatible with the supermanifold grading,
that is, a variable of even degree is commutative and
a variable of odd degree is anticommutative.
By definition, the structure sheaf of $\calM$ is locally isomorphic to
$C^{\infty}(U) \otimes S^{\bullet}(V)$,
where
$U$ is a local chart on $M$,
$V$ is a graded vector space, and $S^{\bullet}(V)$ is a free
graded commutative ring on $V$.
For rigorous mathematical definitions, we refer to 
\cite{Carmeli,Varadarajan}.

In this paper, we only consider nonnegatively graded manifolds.
An \textit{N-manifold} (i.e.,~a nonnegatively graded manifold) $\calM$
equipped with a graded symplectic structure
$\omega$ of degree $n$ is called {\it P-manifold} of degree $n$ and
denoted by $({\calM},\omega)$.
$\omega$ is also called $P$-structure.
We denote the degree of a function $f\in C^\infty({\calM})$ by $|f|$.
A graded Poisson bracket on $C^\infty ({\calM})$ is defined as
$    \{f,g\} = (-1)^{|f|+n+1} i_{X_f} i_{X_g}\omega$,
where
the Hamiltonian vector field $X_f$ is defined by the equation
$i_{X_f}\omega= - \delta f$
for any $f\in C^\infty({\calM})$, where
$\delta$ is the differential on $\calM$.
A vector field $Q$ on $\calM$ is called homological if $Q^2=0$.
\begin{definition}
A \textsl{QP-manifold} is a $P$-manifold $(\calM,\omega)$ endowed with a degree 1 homological vector field $Q$ such that $L_Q \omega =0$
\cite{Schwarz:1992nx}.
\end{definition}
We call the homological vector field $Q$ a Q-structure and
the corresponding triple $(\calM,\omega,Q)$
a QP-manifold.
For any QP-manifold, there exists
a Hamiltonian function $\Theta\in C^{\infty}(\calM)$ 
associated to the homological vector field $Q$
with respect to the graded Poisson bracket $\{-,-\}$, that is,
\beq
Q=\{\Theta,-\}.\label{hamiltonianvector}
\eeq
Then the homological condition, $Q^2=0$, implies that
$\Theta$ is a solution of the \textit{classical master equation}
\begin{equation}
\{\Theta,\Theta\}=0.
\label{cmaseq}
\end{equation}
Such $\Theta$ is called a homological function or a Hamiltonian function.
In order to construct a QP-manifold, 
we specify a symplectic structure $\omega$ and 
a Hamiltonian function $\Theta$. 
The classical master equation provides one unique geometric condition
on the objects involved.
In the next subsection,
we demonstrate how the structures of a Courant algebroid 
emerge from a solution of the classical master equation 
in QP-manifold of degree $2$.

\subsection{QP-manifolds of degree $2$}
\noindent
In this paper, we want to formulate a specific type of Courant algebroid, 
the Poisson Courant algebroid, using the above-mentioned supergeometric method.
This formulation is possible due to the well-known equivalence
of 
a QP-structure of degree $2$ with a 
Courant algebroid on a vector bundle $E$
 \cite{Roytenberg99}. 
This subsection reviews the construction of the 
Courant algebroid on a general vector bundle $E$ including 
the special case where $E=TM\oplus T^*M$. 
In the next section, the construction is applied to the
Poisson Courant algebroid and its properties are discussed.

First, we start with  a vector bundle $E$ over a smooth manifold $M$
with fiber $V$
and a graded manifold $\calM = T^*[2]E[1]$.
Here $E[n]$ denotes the shift of fiber degree by $n$. 

Let $x^i$ be a local coordinate on $M$ and $e^a$ a basis of sections of $E$.
Local coordinates on the graded double vector bundle $\calM$ are denoted 
by $(x^i, \qone^a, \xi_i)$ with
degrees $(0,1,2)$. 
Local coordinates on $E[1]$ are $(x^i, \qone^a)$,
where $\qone^a$ is a local coordinate on fiber $V[1]$ of $E[1]$.
By assuming a fiber metric $\bracket{-}{-}$, we identify
$V^*[1]$ with $V[1]$. Therefore, we use the same local coordinate $\qone^a$ 
for the cotangent space $T^*[2]$ of the fiber of $V[1]$.
Finally, $\xi_i$ is a local coordinate on the fiber of $T^*[2]M$. 
The structure of the graded manifold can schematically 
be represented by the following diagram,
\begin{eqnarray}
\begin{CD}
\calM @>  >> E[1] \\
@V  VV @V  VV \\
T^*[2]M @>  >> M
\end{CD}
\nonumber
\end{eqnarray}

We consider the canonical embedding map of the vector bundle $E$ 
into $\calM$:
\[
j: E \oplus TM \rightarrow \calM~.\label{canonicalembedding}
\] 
The embedding map $j$ can be written using local coordinates by
\begin{equation}
	j: \left(x^i, e^a, \frac{\partial}{\partial x^i}\right) 
\mapsto (x^i, \qone^a, \xi_i). \notag
\end{equation}
For a section $e \in \Gamma (E)$ the pushforward is a function,
$j_* e \in C^{\infty}(E[1])$.
We use the same symbol for $E$ and $jE$, if there is no risk of confusion.

We decompose the structure sheaf by degree,
i.e., the space of functions on $\calM$ as
$C^{\infty}(\calM)  = \sum_{i \geq 0} C_i(\calM)$,
where $C_i(\calM)$ is the space of smooth functions of degree $i$.
We have the following equivalences by the map $j$:
\begin{align}
& C_0(\calM) \simeq C^{\infty}(M), \notag \\
& C_1(\calM) \simeq \Gamma(E), \notag \\
& C_2(\calM) \simeq \Gamma (\wedge^2 E \oplus TM)
\notag \\
& etc. 
\notag
\end{align}

The next step is to introduce a graded symplectic form 
of degree $2$ on $\calM$. We take the following symplectic structure,
\begin{eqnarray}
&& \omega = \delta x^i \wedge \delta \xi_i
+ \frac{1}{2} k_{ab} \delta \qone^a \wedge \delta \qone^b,
\label{QP2symplectic}
\end{eqnarray}
where $\bracket{\qone^a}{\qone^b} = k^{ab}$ is the fiber metric.
This defines a P-structure and leads to the corresponding
graded Poisson bracket
$\sbv{x^i}{\xi_j} = \delta^i{}_j$ and
$\sbv{\qone^a}{\qone^b} = k^{ab}$.

Finally, a Q-structure on $\calM$
is defined by a homological function $\Theta$ of degree $3$
as in \eeqref{hamiltonianvector}.
Using local coordinates, 
the general form of a degree $3$ function is given by
\begin{eqnarray}\label{QCourant}
&& \Theta = \rho^i{}_a(x) \xi_i \qone^a 
+ \frac{1}{3!} C_{abc}(x) \qone^a \qone^b \qone^c,
\end{eqnarray}
where $\rho^i{}_a(x)$ and $C_{abc}(x)$ are arbitrary local functions of $x$.
The homological function satisfies the classical master equation, 
$\{\Theta,\Theta\}=0$. 
This gives a set of relations among the degree zero local functions 
$\rho^i{}_a(x)$ and $C_{abc}(x)$. 
The above triple $({\cal M},\omega,Q)$ defines a QP-manifold of degree $2$. 

The operations on the Courant algebroid are defined using 
a graded Poisson bracket and \textit{derived brackets}. 
The pseudo-metric, Dorfman bracket and anchor map are then 
reconstructed via the following expressions:
\begin{align}
\bracket{e^1}{e^2} &\equiv j^* \sbv{j_* e^1}{j_* e^2}, \notag \\
\courant{e^1}{e^2} &\equiv - j^* \sbv{\sbv{j_* e^1}{\Theta}}{j_* e^2}, 
\notag \\
\rho(e) f &\equiv  j^* \sbv{j_* e}{\sbv{\Theta}{j_* f}},
\label{CourantStructure}
\end{align}
where $f$ is a function on $M$ and $e, e^1, e^2 \in \Gamma (E)$. 
As a consequence of the classical master equation,
these three operations 
satisfy the defining relations of a Courant algebroid.

In the case of $E = TM \oplus T^*M$,
we take local coordinates $(x^i, q^i, p_i, \xi_i)$
with degree $(0,1,1,2)$,
where $x^i$ are local coordinates on $M$,
$q^i$ is a local basis on the fiber of $T[1]M$,
$p_i$ is a local basis on the fiber of $T^*[1]M$
and $\xi_i$ are local coordinates on the fiber of $T^*[2]M$. 
Note that we have identified $\qone^a = (q^i, p_j)$.
The fiber metric is taken as
$k =
\left(
\begin{matrix}
0 & \delta^i{}_j \\
\delta_j{}^i & 0
\end{matrix}
\right)
$. 
Then the graded symplectic form is given by
\begin{eqnarray}
&& \omega = \delta x^i \wedge \delta \xi_i
+ \delta q^i \wedge \delta p_i.\label{QP2symplectic1}
\end{eqnarray}
If we take the Q-structure function as
\begin{eqnarray}
&& \Theta = \xi_i q^i + \frac{1}{3!} H_{ijk}(x) q^i q^j q^k,
\label{standardCATheta}
\end{eqnarray}
we get the standard Courant algebroid by \eeqref{CourantStructure}.
The anchor $\rho$ becomes the natural projection 
from $TM \oplus T^*M$ to $TM$
and the Dorfman bracket becomes 
\eeqref{standardDorfmanbracket}.

\section{{Supergeometric description of 
Poisson Courant algebroids}}\label{PCAfromQP}
\noindent
In this section, we reformulate the Poisson Courant algebroid, 
the Courant algebroid on a Poisson manifold
in \cite{Asakawa:2014kua}, using supergeometry.
First, we give its definition and then 
construct the corresponding QP-manifold.

\begin{definition}
Let $(M, \pi)$ be a Poisson manifold with a Poisson structure
$\pi \in \Gamma (\wedge^2 TM)$ and
$R \in \Gamma(\wedge^3 TM)$ be a $3$-vector field, 
which is closed with respect to 
the Poisson bivector field, $[\pi, R]_S=0$,
where $[-,-]_S$ is the Schouten bracket on multivector fields on 
$\wedge^{\bullet} TM$. 

A \textsl{Poisson Courant algebroid} 
is a vector bundle 
$E = TM \oplus T^*M$ over the Poisson manifold $M$, 
which incorporates the three operations of a Courant algebroid.
The inner product $\langle -, - \rangle$ on $TM \oplus T^*M$
is the same as in the standard Courant algebroid case.
The bundle map $\rho: TM \oplus T^*M \rightarrow TM$
is defined by $\rho(X+\alpha) = \pi^{\sharp}(\alpha)$,
where $\pi^{\sharp}:T^*M \rightarrow TM$
\footnote{
$\pi^{\sharp}: T^*M \rightarrow TM$
is defined by the map
$\pi^{\sharp}(\alpha)
= \pi^{ij} \alpha_i(x) \frac{\partial}{\partial x^j}
$ for any $1$-form, $\alpha = \alpha_i(x) dx^i$.
}.
The bilinear operation
is defined by
\begin{equation}
\courantr{X+\alpha}{Y+\beta} \equiv [\alpha, \beta]_{\pi} + 
L^{\pi}_{\alpha}Y 
-\iota_{\beta} d_{\pi} X - \iota_{\alpha} \iota_{\beta} R, \notag
\end{equation}
where
$X + \alpha, Y + \beta \in \Gamma (TM \oplus T^*M)$,
$d_{\pi}(-) = [\pi, -]_S$
and $[-, -]_{\pi}: T^*M \times T^*M \rightarrow T^*M$
is the Koszul bracket given by 
$[\alpha, \beta]_{\pi} = 
L_{\pi^{\sharp} (\alpha)}\beta
- L_{\pi^{\sharp} (\beta)} \alpha 
- d(\pi(\alpha, \beta))$.
The data of a Poisson Courant algebroid can then be encoded in the quadruple $(E = TM \oplus T^*M, 
\langle -, - \rangle,
\courantr{-}{-},
\rho = 0 \oplus \pi^{\sharp}
)$. 
\end{definition}

We can regard the Poisson Courant algebroid as
a contravariant object associated to the standard Courant algebroid.
Contravariant geometry is a differential calculus in which the roles of
$TM$ and $T^*M$ are exchanged
\cite{Fernandes00,Vaisman}.
Therefore, 
we call $\courantr{-}{-}$ the contravariant Dorfman bracket 
and
we can call this structure 
the \textsl{contravariant Courant algebroid}.

After giving the definition of the Poisson Courant algebroid, we reconstruct this algebroid by supergeometric methods. For this, we use the same graded manifold $\calM = T^*[2]T^*[1]M$ as in the case of the standard Courant algebroid. We also take the same symbols for the local coordinates $(x^i, q^i, p_i, \xi_i )$ and the canonical graded symplectic form
(\ref{QP2symplectic1}).

Then the homological function defining the Q-structure for the Poisson Courant algebroid is 
\begin{eqnarray}
\Theta
=
\pi^{ij}(x) \xi_i p_j
- \frac{1}{2} \frac{\partial \pi^{jk}}{\partial x^i}(x) q^i p_j p_k
+ \frac{1}{3!} R^{ijk}(x) p_i p_j p_k,
\label{PoissonCATheta}
\end{eqnarray}
where
$\pi = \frac{1}{2} \pi^{ij}(x)
\frac{\partial}{\partial x^i} \wedge \frac{\partial}{\partial x^j}
\in \Gamma (\wedge^2 TM)$ is a bivector field and
$R = \frac{1}{3!} R^{ijk}(x) \frac{\partial}{\partial x^i}
\wedge \frac{\partial}{\partial x^j}
\wedge \frac{\partial}{\partial x^k}$ is a $3$-vector field. 
The function $\Theta$ satisfies $\sbv{\Theta}{\Theta} =0$,
i.e., it defines a Q-structure,
if and only if $\pi$ is a Poisson bivector field
and $R$ satisfies $[\pi, R]_S=0$, i.e., it is 
$d_{\pi}$ closed. 
These two conditions 
are exactly the ones required for the Poisson Courant algebroid. 

The derived brackets define all the operations on 
$C^{\infty}(T^*[1]M \oplus T[1]M)
\cong \Gamma (TM \oplus T^*M)$ appearing in the Poisson Courant algebroid as
\begin{eqnarray}
\rho(X+\alpha) f(x)
&=& j^{*}\sbv{\sbv{X^i(x) p_i + \alpha_i(x) q^i}{\Theta}}{f(x)},
\\~
\courantr{X+ \alpha}{Y + \beta}
&=&
j^{*} \sbv{\sbv{X^{i}(x) p_i + \alpha_{i}(x) q^i}{\Theta}}
{Y^{j}(x) p_j + \beta_{j}(x) q^j},
\\
\langle X + \alpha, Y + \beta \rangle &=&
j^{*} \sbv{j_* (X + \alpha)}{j_* (Y + \beta)}.
\end{eqnarray}
We can confirm the above relations using local coordinates.
The first equation gives
\begin{eqnarray}
\rho(X+\alpha) f(x) = j^{*}\sbv{\sbv{X^i(x) p_i + \alpha_i(x) q^i}{\Theta}}{f(x)}
= \pi^{ij} \alpha_i \frac{\partial f}{\partial x^j}(x),
\end{eqnarray}
and thus this is the anchor map
$\rho = 0 \oplus \pi^{\sharp}:TM \oplus T^*M \rightarrow TM
$.
The second equation gives
\begin{eqnarray}
\courantr{X + \alpha}{Y + \beta}
&=&
j^{*} \sbv{\sbv{X^{i}(x) p_i + \alpha_{i}(x) q^i}{\Theta}}
{Y^{j}(x) p_j + \beta_{j}(x) q^j}
\nonumber\\
&=&
\left(
\alpha_j \pi^{jk} \frac{\partial Y^i}{\partial x^k}
+ \frac{\partial X^i}{\partial x^j} \pi^{jk} \beta_k
- \frac{\partial X^j}{\partial x^k} \pi^{ki} \beta_j
- \frac{\partial \alpha_j}{\partial x^k} \pi^{ki} Y^j
\right.
\nonumber \\ &&
\left.
+ \frac{\partial \pi^{ki}}{\partial x^j} X^j \beta_k
- \frac{\partial \pi^{ji}}{\partial x^k} \alpha_j Y^k
- R^{jki} \alpha_j \beta_k
\right) j^{*} p_i
\nonumber \\ &&
+ \left(
\alpha_j \pi^{jk} \frac{\partial \beta_i}{\partial x^k}
+ \frac{\partial \alpha_i}{\partial x^j} \pi^{jk} \beta_k
+ \frac{\partial \pi^{jk}}{\partial x^i} \alpha_j \beta_k
\right) j^{*} q^i
\nonumber \\
&=& [\alpha, \beta]_{\pi} +
L^{\pi}_{\alpha}Y - \iota_{\beta} d_{\pi} X
- R(\alpha, \beta, -),
\end{eqnarray}
and thus is the Dorfman bracket of the Poisson Courant algebroid 
on 
$TM \oplus T^*M$.
The third equation is the same as in the case of the standard Courant algebroid.

The classical master equation, $\sbv{\Theta}{\Theta} =0$, leads to 
the following conditions on these operations,
\begin{eqnarray}
&& \rho (\courantr{e_1}{e_2}) =[\rho(e_1), \rho(e_2)],
\\
&& \courantr{e_1}{\courantr{e_2}{e_3}}
=\courantr{\courantr{e_1}{e_2}}{e_3}
+\courantr{e_2}{\courantr{e_1}{e_3}},
\\
&& \rho (e_1)\bracket{e_2}{e_3}
=\bracket{\courantr{e_1}{e_2}}{e_3}+\bracket{e_2}{\courantr{e_1}{e_3}},
\\
&& \rho
(e_1)
\langle e_2,e_3\rangle
=\langle e_1, \courantr{e_2}{e_3} +\courantr{e_3}{e_2} \rangle,
\end{eqnarray}
\noindent 
where $e_i \in \Gamma (TM \oplus T^*M)$. 
These are the relations required for the Poisson Courant algebroid.

Thus, in the graded manifold method, the difference between the standard Courant algebroid and
the Poisson Courant algebroid
lies in the choice of the homological function. 
The most general homological function $\Theta$ 
on $E = TM\oplus T^*M$ possible for the Courant algebroid is
\begin{equation}\label{GeneralHomologicalFunction}
\Theta = \tau^i{}_j(x)\xi_i q^j + \sigma^{ij}(x)\xi_i p_j 
+ \frac{1}{3!}H_{ijk}(x)q^i q^j q^k 
+ \frac{1}{2}F_{ij}{}^{k}(x)q^i q^j p_k 
+ \frac{1}{2}Q_{i}{}^{jk}(x)q^i p_j p_k 
+ \frac{1}{3!}R^{ijk}(x)p_i p_j p_k.
\end{equation}
The classical master equation then imposes structural 
restrictions onto the expansion coefficients. 
One of the conditions for $\tau$ and $\sigma$ is
$\tau{}^i{}_k \sigma{}^{jk} + \sigma{}^{ik} \tau{}^j{}_k = 0$.
The two simplest solutions are $\tau =0$, $\sigma \neq 0$ or 
$\tau \neq 0$, $\sigma =0$.
In 
the standard Courant algebroid
case,
$\tau^i{}_j = \delta^i{}_j$ and
$\sigma = F = Q = R = 0$,
and 
in
the Poisson Courant algebroid
case, 
$\sigma = \pi$, 
$Q_{i}{}^{jk}(x) = -\frac{\partial \pi^{jk}}{\partial x^i}(x)$ 
and $\tau = H = F = 0$.

\section{Duality between $H$-flux and $R$-flux}
\noindent
In this section, we study the meaning of 
$H$-flux geometry and $R$-flux geometry.
We analyze the 'duality' transformation between two Courant algebroids 
in terms of supergeometry
and the homological algebra.
The key operation is a canonical transformation on the graded
symplectic manifold (the P-manifold).
This duality is a generalization 
of the correspondence between de Rham cohomology on differential forms
and Poisson cohomology on multivector fields.

In this section, we denote the homological function $\Theta$ 
of the standard Courant algebroid in \eeqref{standardCATheta}
as $\Theta_H$ and the one of the Poisson Courant algebroid in 
\eeqref{PoissonCATheta} as $\Theta_R$.

\subsection{Flux duality transformations
as canonical transformations}
\label{fluxduality}
\noindent
Suppose the Poisson structure $\pi$ is nondegenerate.
We construct the duality transformation 
between the standard Courant algebroid and the Poisson Courant algebroid,
which is derived from the transformation between the two homological functions 
$\Theta_H$ and $\Theta_R$.
This leads to a duality between the
standard Courant algebroid cohomology
and the Poisson Courant algebroid cohomology.

First,
we define a {\sl canonical transformation} on a P-manifold.
Let $\alpha \in C^{\infty}(\calM)$. 
$e^{\delta_\alpha}$ is the exponential adjoint operation,
$$
e^{\delta_\alpha} f
= f + \sbv{f}{\alpha}
+ \frac{1}{2} \sbv{\sbv{f}{\alpha}}{\alpha}
+ \cdots,
$$
for any $f\in C^{\infty}(\cal M)$.
If $\alpha$ is of degree $n$, 
this transformation preserves degree and satisfies
$\sbv{e^{\delta_\alpha} f}{e^{\delta_\alpha} g} 
= e^{\delta_\alpha} \sbv{f}{g}$, where $f, g \in C^{\infty}(\calM)$.
\begin{definition}
For any function $\alpha$ of degree $n$, 
$e^{\delta_\alpha}$ is called a {\sl canonical transformation}. 
\end{definition}
$e^{\delta_\alpha}$ is also called {\sl twisting} 
\cite{Roytenberg:2001am}.

Both the standard Courant algebroid and the Poisson Courant algebroid 
are realized on the same P-manifold $(T^*[2]T^*[1]M, \omega)$.
Therefore, the duality transformation $T$ from $H$-flux to $R$-flux is 
a symplectomorphism on $T^*[2]T^*[1]M$ such that the two homological functions
are mapped,
$T:\Theta_H \mapsto \Theta_R$,
where
\begin{eqnarray}
\Theta_H &=& \xi_i q^i + \frac{1}{3!} H_{ijk}(x) q^i q^j q^k,
\\
\Theta_R &=& 
\pi^{ij}(x) \xi_i p_j
- \frac{1}{2} \frac{\partial \pi^{jk}}{\partial x^i}(x) q^i p_j p_k
+ \frac{1}{3!} R^{ijk}(x) p_i p_j p_k.
\end{eqnarray}
We denote
$\alpha_p = \frac{1}{2} \pi^{ij}(x) p_i p_j$
and 
$\alpha_q = \frac{1}{2} \pi^{-1}_{ij}(x) q^i q^j$.
$\alpha_p$ generates a so-called $\beta$-transformation and 
$\alpha_q$ generates a $b$-transformation.
{
Note that $\alpha_q$ is a trivial transformation on $\Theta_H$,
$e^{\delta_{\alpha_q}}\Theta_H=\Theta_H$,
and 
$\alpha_p$ is a trivial transformation on $\Theta_R$,
$e^{\delta_{\alpha_p}}\Theta_R=\Theta_R$,
from $\sbv{\Theta_H}{\alpha_q}=0$ and
$\sbv{\Theta_R}{\alpha_p}=0$.}
By direct computation,
we get the relation
{
\begin{eqnarray}
\Theta_R = e^{\delta_{\alpha_p}} 
e^{-\delta_{\alpha_q}} e^{\delta_{\alpha_p}} \Theta_H,
\label{twistedHRfluxTheta}
\end{eqnarray}
}
where $R = \wedge^3 \pi^{\sharp} H$.
On the basis of the QP-manifold, this canonical transformation acts as
{
\begin{eqnarray}
&& e^{\delta_{\alpha_p}} e^{- \delta_{\alpha_q}} e^{\delta_{\alpha_p}} x^i = x^i,
\label{HtoRcanonical1}
\\
&& e^{\delta_{\alpha_p}} e^{- \delta_{\alpha_q}} e^{\delta_{\alpha_p}} q^i = \pi^{ij}(x) p_j,
\label{HtoRcanonical2}
\\
&& e^{\delta_{\alpha_p}} e^{- \delta_{\alpha_q}} e^{\delta_{\alpha_p}} p_i 
= - \pi^{-1}_{ij}(x) q^j,
\label{HtoRcanonical3}
\\
&& e^{\delta_{\alpha_p}} e^{- \delta_{\alpha_q}} e^{\delta_{\alpha_p}} \xi_i 
= \xi_i 
+ \frac{\partial \pi^{jk}}{\partial x^i} \pi^{-1}_{kl}(x) p_j q^l.
\label{HtoRcanonical4}
\end{eqnarray}
The Liouville $1$-form is transformed as
\begin{eqnarray}
&& \vartheta
= \xi_i \delta x^i - q^i \delta p_i
\rightarrow  \xi_i \delta x^i - p_i \delta q^i 
= \vartheta^{\prime}.
\end{eqnarray}
}
Note that 
{$j^* \pi^{\sharp} (\alpha) 
= e^{\delta_{\alpha_p}} 
e^{-\delta_{\alpha_q}} e^{\delta_{\alpha_p}}j^* \alpha$}
for a $1$-form $\alpha$.

The standard Courant algebroid with $H$-flux and 
the Poisson Courant algebroid with $R$-flux are transformed
by a symplectomorphism between two QP-manifolds
if $\pi$ is nondegenerate.
From this observation,
we understand
%
that a flux duality is characterized by 
a symplectomorphism between two QP-manifolds
$T:(\calM, \omega, Q) \rightarrow (\calM, \omega, Q^{\prime})$.
In this sense, $B$- and $\beta$-transformations are 
special cases of flux duality transformations.

The flux duality transformation in \eeqref{twistedHRfluxTheta} 
will be reinterpreted in 
the sigma model language on an open manifold
as a change of boundary conditions in subsection \ref{CSMduality}.

\subsection{Two cohomologies on $T^*[2]T^*[1]M$}
\noindent
$Q_H = \sbv{\Theta_H}{-}$ and $Q_R = \sbv{\Theta_R}{-}$ are two coboundary 
operators increasing degree by $1$ on the space of functions on the 
QP-manifold $\calM = T^*[2]T^*[1]M$.
Expanding $C^{\infty}(\calM) = \sum_{i \geq 0} C_i(\calM)$ by degree,
$(C_{\bullet}(\calM), Q)$ becomes the complex for both coboundary operators,
the so-called standard complex of 
the Courant algebroid \cite{Roytenberg:2002nu},
where $Q = Q_H$ or $Q = Q_R$.
$(C_{\bullet}(\calM), Q_H)$ defines the standard Courant algebroid
cohomology 
$H_{SCA}^{\bullet}(M, Q_H)$ and 
$(C_{\bullet}(\calM), Q_R)$ defines the Poisson Courant algebroid
cohomology 
$H_{PCA}^{\bullet}(M, Q_R)$.

Both cohomologies are known cohomologies
on the special subspaces.
As discussed in section \ref{CAandsupergeometry},
there exists the embedding map of the Courant algebroid 
to a graded manifold, 
$j: E \oplus TM
\rightarrow T^*[2]T^*[1]M$,
where $E = TM \oplus T^*M$ is the Courant algebroid.
We take a local basis 
$\left(\frac{\partial}{\partial x^i}, dx^i, 
\frac{\partial}{\partial x^i} \right)$
on $E \oplus TM$
\footnote{The first $\frac{\partial}{\partial x^i}$ is the
basis of the tangent bundle in the Courant algebroid
and the third $\frac{\partial}{\partial x^i}$ is the
basis of the tangent bundle of the image of the anchor map.}.
The local basis is mapped as
$j: \left(\frac{\partial}{\partial x^i}, dx^i, 
\frac{\partial}{\partial x^i} \right)
\mapsto (p_i, q^i, \xi_i)$.

First, we consider the standard Courant algebroid with H-flux.
Since $T[1]M$ is isomorphic to $T^*M$,
let us consider the subspace $C^{\infty}(T[1]M)$.
An element $\gamma \in C^{\infty}(T[1]M)$ can be written as
\begin{eqnarray}
\gamma = \frac{1}{s!} \gamma_{i_1 \cdots i_s}(x)
q^{i_1} \cdots q^{i_s}.
\end{eqnarray}
$\gamma$ is mapped to a differential form by pullback,
\begin{eqnarray}
j^* \gamma = \frac{1}{s!} \gamma_{i_1 \cdots i_s}(x)
dx^{i_1} \wedge \cdots \wedge dx^{i_s}.
\end{eqnarray}
Thus, $C^{\infty}(T[1]M)$ is equivalent to the space of differential
forms $\Omega^{\bullet}(M)$.
We can easily show that the operation of
$Q_H$ on $C^{\infty}(T[1]M)$
is the de Rham differential $d$ on $\Omega^{\bullet}(M)$,
\begin{eqnarray}
d (j^* \gamma) = - j^* Q_H \gamma.
\end{eqnarray}
Therefore, the restriction of the standard Courant cohomology 
$H_{SCA}^{\bullet}(M, Q_H)$ to 
$C^{\infty}(T[1]M)$ is equivalent to the de Rham cohomology,
\begin{eqnarray}
H_{SCA}^{\bullet}(M, Q_H)\Big|_{C^{\infty}(T[1]M)} \simeq H^{\bullet}_{dR}(M, d).
\end{eqnarray}

Next, we consider the Poisson Courant algebroid case.
Note that $T^*[1]M$ is isomorphic to $TM$.
Let us consider the subspace $C^{\infty}(T^*[1]M)$, 
whose elements can be written as
\begin{eqnarray}
u = \frac{1}{s!} u^{i_1 \cdots i_s}(x)
p_{i_1} \cdots p_{i_s}.
\end{eqnarray}
The pullback maps $u$ to a multivector field,
\begin{eqnarray}
j^* u = \frac{1}{s!} u^{i_1 \cdots i_s}(x)
\frac{\partial}{\partial x^{i_1}} \wedge \cdots \wedge
\frac{\partial}{\partial x^{i_s}}.
\end{eqnarray}
Hence, $C^{\infty}(T^*[1]M)$ is equivalent to the space of multivector 
fields $T_{poly}^{\bullet}(M)$.

If we put $R=0$, then $\Theta_R$ is given by
\begin{eqnarray}
\Theta_R \Big|_{R=0} =
- \sbv{\pi}{\Theta_0}
= - \left\{\frac{1}{2} \pi^{jk}(x) p_j p_k, \xi_i q^i \right\},
\label{PoissonCAR}
\end{eqnarray}
and
\begin{eqnarray}
Q_R u = \sbv{\Theta_R}{u}
= - \sbv{\sbv{\pi}{\Theta_0}}{u},
\label{QRoperation}
\end{eqnarray}
for $u \in C^{\infty}(T^*[1]M)$.
Since the derived bracket 
$\sbv{\sbv{-}{\Theta_0}}{-}$ 
is equivalent to the Schouten bracket $[-,-]_S$,
and the graded Poisson bracket of $R$ with 
elements of 
$C^{\infty}(T^*[1]M)$
is zero,
$Q_R$ is equivalent to the Poisson differential
$d_{\pi} = [\pi,-]_S$
on $T_{poly}^{\bullet}(M)$,
\begin{eqnarray}
d_{\pi} (j^* u) = - j^* Q_R u~.
\end{eqnarray}
The cohomology defined by the coboundary operator $d_{\pi}$ is the 
Poisson cohomology, $H_P^k(M, d_{\pi})$.
Therefore, the restriction of the Poisson Courant cohomology 
$H_{PCA}^{\bullet}(M, Q_R)$ to 
$C^{\infty}(T^*[1]M)$ is equivalent to the Poisson cohomology,
\begin{eqnarray}
H_{PCA}^{\bullet}(M, Q_R)\Big|_{C^{\infty}(T^*[1]M)} \simeq H^{\bullet}_{P}(M, d_{\pi}).
\end{eqnarray}

\subsection{Duality of cohomologies}
\noindent
Since we are considering 
an even dimensional manifold
and a nondegenerate Poisson structure,
we can prove that 
$H^{\bullet}_{SCA}(M, Q_H)$ is isomorphic to $H_{PCA}^{\bullet}(M, Q_R)$.
This is a generalization of 
the well known result that if $M$ is symplectic,
the de Rham cohomology and the Poisson cohomology are
isomorphic, 
$H^k_{dR}(M, d) \simeq H_{P}^k(M, d_{\pi})$ \cite{Vaisman}.

The map 
$\wedge^k \pi^{\sharp}: \Omega^k(M) \rightarrow T_{poly}^k(M)$ induces
a homomorphism between de Rham cohomology and Poisson cohomology,
$$
\wedge^k \pi^{\sharp}: H_{dR}^k(M, d) \rightarrow 
H_P^k(M, d_{\pi}).
$$
If $\pi^{-1}$ is symplectic,
the de Rham cohomology on $\Omega^{\bullet}(M)$ 
is isomorphic to the Poisson cohomology:
\begin{eqnarray}
&& H_{dR}^k(M, d) \simeq H_P^k(M, d_{\pi}).
\label{deRhamPoissonCoho}
\end{eqnarray}

In our duality theory, 
the de Rham cohomology is extended to 
$H_{SCA}^{\bullet}(M, Q_H)$ and the Poisson cohomology is extended to
$H_{PCA}^{\bullet}(M, Q_R)$.
Duality of H-flux and R-flux is understood as 
duality of $H_{SCA}^{\bullet}(M, Q_H)$ and
$H_{PCA}^{\bullet}(M, Q_R)$
on the same space $C^{\infty}(\calM)$.

Let $f$ be a general function on $\calM = T^*[2]T^*[1]M$,
$f \in C^{\infty}(\calM)$.
From \eeqref{twistedHRfluxTheta}, 
it follows that 
$Q_H f = 0 \Leftrightarrow 
Q_R (e^{\delta_{\alpha_p}} 
e^{-\delta_{\alpha_q}} e^{\delta_{\alpha_p}} f) = 0$
and 
$f = Q_H g \Leftrightarrow f = 
Q_R (e^{\delta_{\alpha_p}} 
e^{-\delta_{\alpha_q}} e^{\delta_{\alpha_p}} g)$.
Therefore,
we obtain maps from elements of the $Q_H$-complex 
to elements of the $Q_R$-complex,
\begin{eqnarray}
T: f \mapsto e^{\delta_{\alpha_p}} 
e^{-\delta_{\alpha_q}} e^{\delta_{\alpha_p}} f.
\label{}
\end{eqnarray}
The flux duality map of complexes
$T: C_{\bullet}(\calM) \rightarrow C_{\bullet}(\calM)$
gives rise to the isomorphism of cohomologies,
$T: H_{SCA}^{\bullet}(M, d_H) \rightarrow H_{PCA}^{\bullet}(M, d_R)$.

We obtain the following theorem.
\begin{theorem}
Let $\pi$ be a nondegenerate Poisson structure, that is, 
$\pi^{-1}$ is symplectic, and $R = \wedge^3 \pi^{\sharp} H$.
Then,
the standard Courant algebroid cohomology is isomorphic to 
the Poisson Courant algebroid cohomology,
\begin{eqnarray}
&& H_{SCA}^k(M, Q_H) \overset{\sim}{\rightarrow} H_{PCA}^k(M, Q_R).
\end{eqnarray}
\end{theorem}

\section{{Poisson Courant algebroids from double field theory}}
\label{PCAfromDG}
\noindent
In this section, we show that 
the Poisson Courant algebroid is a 
solution of the section condition (the strong constraint) in 
double field theory.
This shows that the Poisson Courant algebroid is 
directly connected to the geometry of double field theory.

\subsection{{Supergeometric formulation, Poisson structure 
and double field theory}}
\noindent
We start with the supergeometric formulation of the
geometry of double field theory \cite{Deser:2014mxa}.
We take a doubled configuration space $\widehat{M}$ in $2d$ dimensions
with local coordinates $(y^i, \tilde{y}_i)$
and a QP-manifold of degree $2$, 
$T^*[2]\widehat{M}$ with fiber coordinates,
$(\eta_i, \tilde{\eta}^i)$,
such that $\sbv{y^i}{\eta_j} = \sbv{\tilde{y}_j}{\tilde{\eta}^i} 
= \delta^i{}_j$.
Moreover, we introduce degree one canonical conjugate coordinates 
$(q^i, p_i)$ such that $\sbv{q^i}{p_j} = \delta^i{}_j$.

On this P-manifold,
the geometry of double field theory is formulated using the Q-structure
homological function,
\begin{eqnarray}
\Theta_C = \eta_i q^i + \tilde{\eta}^i p_i. \label{DFTnoTwist}
\end{eqnarray}
The classical master equation,
$\sbv{\Theta_C}{\Theta_C} =0$,
gives rise to the section condition,
\begin{eqnarray}
\tilde{\eta}^i \eta_i =0.
\label{sectioncondition}
\end{eqnarray}
The C-bracket is constructed by the derived bracket,
\begin{eqnarray}
[e_1, e_2]_C = -\frac{1}{2}[j^* \sbv{\sbv{j_* e_1}{\Theta_C}}{j_* e_2} - (1\leftrightarrow 2)],
\end{eqnarray}
where $e_1, e_2$ are sections of $T\widehat{M} \oplus T^*\widehat{M}$.

We choose a nontrivial physical configuration space, 
a $d$-dimensional submanifold $M \subset \widehat{M}$
with local coordinate $x^i$
under the assumption that $M$ has a Poisson structure $\pi$.
We can consider a local coordinate transformation
with following Jacobian,
\begin{eqnarray}
\frac{\partial(x, \tilde{x})}{\partial (y, \tilde{y})} 
=
\left(
\begin{matrix}
\frac{\partial {x}^i}{\partial y^j} 
& \frac{\partial {x}^i}{\partial \tilde{y}_j} 
\\
\frac{\partial \tilde{x}_i}{\partial y^j} 
& \frac{\partial \tilde{x}_i}{\partial \tilde{y}_j} 
\end{matrix}
\right)
= 
\left(
\begin{matrix}
\delta^i{}_j 
& \pi^{ij}  \\
0 & 
\delta_i{}^j 
\end{matrix}
\right).
\label{diffeofromytox}
\end{eqnarray}
Alternatively, this local coordinate transformation
can be realized as a twist of the original $\Theta_C$
by a canonical function 
$\alpha_{p} = \frac{1}{2} \pi^{ij}(x) p_i p_j$.
Here, we denote the original homological function as 
$\Theta_C^{\prime} = \xi_i q^i + \tilde{\xi}^i p_i$.
The canonical transformation deforms the homological function,
\begin{eqnarray}
\Theta_C = e^{\alpha_p} \Theta_C^{\prime} &=& \xi_i q^i 
+ \tilde{\xi}^i p_i 
+ \pi^{ij} \xi_i p_j
- \frac{1}{2} \frac{\partial \pi^{jk}}{\partial x^i}(x) q^i p_j p_k.
\end{eqnarray}
This corresponds to the change of variables,
\begin{eqnarray}
\eta_i &=& \xi_i
- \frac{1}{2} \frac{\partial \pi^{jk}}{\partial x^i}(x) p_j p_k,
\label{xietatransformation}
\\
\tilde{\eta}^i &=& \tilde{\xi}^i 
- \pi^{ij} \xi_j.
\end{eqnarray}
We need the second term in \eeqref{xietatransformation}
for consistency 
of the Poisson structure with the local coordinate transformation
\eeqref{diffeofromytox}.
%
The section condition is deformed to
\begin{eqnarray}
\tilde{\xi}^i \left(\xi_i 
- \frac{1}{2} \frac{\partial \pi^{jk}}{\partial x^i}(x) p_j p_k
\right)
=0.
\label{sectioncondition2}
\end{eqnarray}
Finally, we take 
$\tilde{\xi}^i=0$ corresponding to the submanifold defined by 
$\tilde{x}_i = 0$,
and obtain homological functions 
of both standard and Poisson Courant algebroids,
\begin{eqnarray}
\Theta_C|_{\tilde{x}=0} &=& \Theta_{H=0}
+ \Theta_{R=0}.
\end{eqnarray}
Note that $\Theta_C$ defines a double complex, since $\sbv{\Theta_{H=0}}{\Theta_{R=0}}=0$.
In this paper, we analyze these two Courant algebroids.
In fact, $\Theta_{H=0} + \Theta_{R=0}$
defines a Lie bialgebroid on $TM \oplus T^*M$.
\footnote{In the case where $H$ and $R$ are nonzero this defines 
a proto-Lie bialgebroid \cite{Roytenberg:2002nu, Kos2}.
Solutions of the classical master equation 
give a dependency between both fluxes if the Courant algebroid is exact
\cite{Severa}.}
Since we change the section condition $\tilde{\eta}^i=0$ to $\tilde{\xi}^i=0$,
in general, the configuration space $M$ is not embedded as
a direct product $M \times \tilde{M}$ in the doubled space,
but is a nontrivial submanifold of the doubled configuration space.

\subsection{{Poisson Courant algebroid $R$-flux in double field theory}}
\noindent
Finally, we want to discuss how the $R$-flux of the Poisson Courant algebroid relates to the $R$-flux in double field theory. For this, we compute the $B$- and $\beta$-twist of the homological function \eeqref{DFTnoTwist}. This gives the following description of all fluxes $H$, $F$, $Q$ and $R$ in terms of their potentials
\begin{align}
	e^{-\delta_\beta}e^{-\delta_B} \Theta_C &= (\eta_i - B_{mi}\teta^m)q^i + (\teta^i - \eta_m\beta^{mi} + \teta^n B_{nm}\beta^{mi})p_{i} \notag \\
	&\qquad + \frac{1}{2}\left[-B_{in}\tpar^i B_{rs} + \partial_n B_{rs}\right] q^n q^r q^s \notag \\
	&\qquad + \left[\frac{1}{2}\tpar^i B_{mn} + (B_{lm}\tpar^l B_{ns} - \partial_m B_{ns} + \frac{1}{2} B_{ls}\tpar^{l}B_{mn} - \frac{1}{2}\partial_s B_{mn})\beta^{si}\right]p_{i}q^m q^n \notag \\
	&\qquad + \left[ \frac{1}{2} \partial_i \beta^{hk} - \frac{1}{2}B_{li}\tpar^l \beta^{hk} + \tpar^h B_{in}\beta^{nk} \right. \notag \\
	&\qquad - \left.  \frac{1}{2}\left[- B_{li}\tpar^l B_{rs} + \partial_{i}B_{rs} - B_{ls}\tpar^l B_{ir} + \partial_s B_{ir} + B_{lr}\tpar^l B_{is} - \partial_r B_{is}\right]\beta^{sh}\beta^{rk}\right]q^{i}p_h p_k \notag \\
	&\qquad + \left[\frac{1}{2}\tpar^i \beta^{hk} - \frac{1}{4}\partial_l \beta^{ih}\beta^{lk} - \frac{1}{4}\beta^{li}\partial_l \beta^{hk} + \frac{1}{4}B_{ln}\tpar^l \beta^{ih}\beta^{nk} \right. \notag \\
	&\qquad + \frac{1}{4}B_{ln}\beta^{ni}\tpar^l \beta^{hk} - \frac{1}{2}\tpar^i B_{mn}\beta^{nh}\beta^{mk} \notag \\
	&\qquad \left. + \frac{1}{3!}(-B_{ln}\tpar^l B_{rs} + \partial_{n}B_{rs} - B_{ls}\tpar^l B_{nr} + \partial_s B_{nr} + B_{lr}\tpar^l B_{ns} - \partial_r B_{ns})\beta^{si}\beta^{rh}\beta^{nk}\right] p_{i}p_{h}p_{k}, \label{DFThom}
\end{align}
where $B = \frac{1}{2}B_{ij}(y,\ty)q^i q^j$ and 
$\beta = \frac{1}{2}\beta^{ij}(y,\ty)p_i p_j$ are functions 
on the ordinary coordinates $y$ and their duals $\ty$. 
The Q-structure function of the $H$-twisted standard Courant algebroid 
\eeqref{standardCATheta} realizes the double field theory $H$-flux 
in the supergravity limit $(\teta^i = \tpar^i = 0)$ 
with vanishing $\beta$-field, $\beta^{ij}=0$. 
Indeed, truncation of \eeqref{DFThom} to this frame leads to the correct 
local description of $H$-flux in terms of its potential 
$H_{nrs}=\frac{1}{2}\partial_{[n}B_{rs]}$. 
On the other hand, truncation to the non-geometric frame, 
where $\eta_i = \partial_i=0$ and $B_{ij}=0$, 
leads to the correct description of $R$-flux in terms of its potential 
$R^{ihk}=\frac{1}{2}\tpar^{[i}\beta^{hk]}$. 
Finally, the truncation to the frame such that $\teta^i = \tpar^i=0$ 
and $B_{ij}=0$ gives the correct description of $R$-flux in terms of 
$R^{ihk}=\frac{1}{2}\beta^{[i|l|}\partial_l \beta^{hk]}$. 
This brings us into the position to compare the $R$-flux of 
the Poisson Courant algebroid to the $R$-flux of double field theory. 
For this, we have to distinguish two cases, which will be discussed 
in the following.

The first case concerns the meaning of the transformation 
$\Theta_H \rightarrow \Theta_R$
\eeqref{twistedHRfluxTheta} of the standard Courant algebroid with $H$-flux to 
the Poisson Courant algebroid $R$-flux. 
For this transformation, we
introduce the Poisson bivector field $\pi$.
Since, as described above, the standard Courant algebroid with $H$-flux 
already works in the supergravity frame with zero $\beta$-field by the 
identification $y^i = x^i$ and $\ty_i = \tx_i = 0$, 
the resulting Poisson Courant algebroid also works in the same frame. 
The Poisson tensor $\pi$ is introduced as additional freedom, which is 
not related to the flux potentials $\beta$ and $B$, and the resulting 
Poisson Courant algebroid with $R$-flux serves as a different way of 
representing an $H$-flux background on a Poisson manifold. 
On the other hand, the term
$- \frac{1}{2} \frac{\partial \pi^{ij}}{\partial x^k} q^k p_i p_j$
in \eeqref{PoissonCATheta}
is a so-called Poisson connection with vanishing curvature.
Therefore, it must be distinguished from
a $Q$-flux term in double field theory.

In the second case, if the Poisson Courant algebroid 
is seen as a stand-alone object, we can make contact to 
the double field theory $R$-flux. Through identification of 
\eeqref{DFThom} to \eeqref{PoissonCATheta} we find
\begin{align}
	\pi^{ij}(x)\xi_i &= \teta^j - \eta_m \beta^{mj}(y,\ty) + \teta^n B_{nm}(y,\ty)\beta^{mj}(y,\ty), \\
	0 &= \eta_i - B_{mi}(y,\ty)\teta^{m}, \label{PCAtoDFT2}
\end{align}
which leads to the identification $\pi^{ij}(x)\xi_i = \teta^j$ or $\pi^{ij}(x)\frac{\partial}{\partial x^i} = \frac{\partial}{\partial \ty_j}$
and we can read off how the section condition is solved. 
Integration of this equation leads to
\begin{equation}
	\ty_i = \int \pi_{ij}^{-1}(x)dx^j.
\end{equation}
Since there is no $H$-flux coefficient in the Poisson Courant algebroid, 
we obtain the relation $B_{ij}=0$, which leads to 
$\frac{\partial}{\partial y^i}=0$ due to \eeqref{PCAtoDFT2}. 
The term $- \frac{1}{2} \frac{\partial \pi^{ij}}{\partial x^k} q^k p_i p_j$
in \eeqref{PoissonCATheta}
is not sourced by 
the potentials $B$ or $\beta$, 
but is a Poisson connection arising from the underlying space,
and its origin is different from the $Q$-flux in double field theory.
Finally, the local description of $R$-flux is then given in terms of 
the $\beta$-potential via
\begin{align}
	R^{ihk} &= \frac{1}{2}\frac{\partial}{\partial \ty_{[i}}\beta^{hk]}(y,\ty)\big|_{y=0,\ty=\int \pi^{-1}(x)dx}
\notag \\
 &= \frac{1}{2}\pi^{j[i}(x)\frac{\partial}{\partial x^j}\beta^{hk]}(x)
 \notag \\
	&= \frac{1}{2}[\pi,\beta]_{S}.
\end{align}
{
To summarize, the Poisson Courant algebroid can be interpreted in 
two different ways, depending on the frame chosen in double field theory.
In order to analyze the property of $R$-flux, 
we can use this correspondence, and 
on spacetime with a Poisson structure, some parts of $R$-flux geometry 
can be analyzed as $H$-flux geometry.
}

\section{Topological sigma models}
\noindent
We want to consider field theoretical models with 
Poisson Courant algebroid symmetry. 
Here, we construct a $3$-dimensional AKSZ sigma model, i.e., 
a theory of a topological membrane with
$3$-vector flux $R$,
following the construction of a topological membrane theory
based on the standard Courant algebroid 
\cite{Park:2000au,Ikeda:2001fq}.
For this purpose,
first we shortly review the concept of AKSZ sigma models
\cite{Alexandrov:1995kv,Cattaneo:2001ys,Roytenberg:2006qz}.

\subsection{AKSZ sigma models}
\noindent
Let $(\calX, D, \mu)$ be a differential graded manifold
$\calX$
with a $D$-invariant nondegenerate measure $\mu$,
where
$D$ is a differential on $\calX$.
Let ($\calM, \omega, Q$) be a QP-manifold,
and let
$\Map(\calX, \calM)$ be the space of smooth maps from $\calX$ to $\calM$.
It means that we consider the sigma model on
the worldvolume $\calX$ embedded into the target space $\calM$.

Since
${\rm Diff}(\calX)\times {\rm Diff}(\calM)$
naturally acts on $\Map(\calX, \calM)$,
$D$ and $Q$ induce differentials $\hat{D}$ and $\check{Q}$, respectively,
on $\Map(\calX, \calM)$.
Explicitly,
$\hat{D}(z, f) = D(z) \delta f(z)$
and $\check{Q}(z, f) = Q f(z)$,
for all $z \in \calX$ and $f \in \Map(\calX, \calM)$.

The {\it evaluation map}
${\rm ev}: \calX \times \Map(\calX, \calM) \longrightarrow \calM$
is defined as
${\rm ev}:(z, f) \longmapsto f(z)$,
for any
$z \in \calX$ and $f \in \Map(\calX, \calM)$.
The {\it chain map} on the space of graded differential forms,
$\mu_*: \Omega^{\bullet}(\calX \times \Map(\calX, \calM))
\longrightarrow \Omega^{\bullet}(\Map(\calX, \calM))$, is defined as
$$\mu_* \omega(f)(v_1, \ldots, v_k)
 = \int_{\calX} \mu(z)
 \omega(z, f) (v_1, \ldots, v_k),$$
for a graded differential form $\omega$,
where $v_{i}$ are a vector fields on $\calX$
and
$\int_{\calX} \mu$ is the Berezin integration on $\calX$.
The composition $\mu_* \ev^*: \Omega^{\bullet}(\calM)
\longrightarrow \Omega^{\bullet}(\Map(\calX, \calM))$
is called {\it transgression map}.

A \textsl{P-structure} $\bomega$
on $\Map(\calX, \calM)$ is defined by
\begin{eqnarray*}
\bomega := \mu_* \ev^* \omega.
\end{eqnarray*}
%
Note that $\bomega$ is nondegenerate and closed
since the operation $\mu_* \ev^*$ preserves these properties.
The corresponding graded Poisson bracket on the mapping space
$\Map(\calX, \calM)$ is also denoted by $\sbv{-}{-}$.
We show later that this bracket is the BV bracket $\mbv{-}{-}$ 
or the Poisson bracket $\sbv{-}{-}_{PB}$.

A \textsl{Q-structure} function
$S$ on $\Map(\calX, \calM)$
is constructed as follows.
$S$ consists of two parts $S =S_0+S_1$.
We take a canonical $1$-form (the Liouville $1$-form) $\vartheta$
on $\calM$ such that
$\omega= - \delta \vartheta$ and
define $S_0 = \iota_{\hat{D}} \mu_* {\rm ev}^* \vartheta$.
Moreover, we define $S_1 = \mu_* \ev^* \Theta$,
where $\Theta$ is the homological function on $\calM$.
Then we can prove that $S$ is a homological function
on $\Map(\calX,  \calM)$, 
\begin{eqnarray}
\sbv{S}{S} =0,
\label{classicalmaster}
\end{eqnarray}
using the definitions of $S_0$ and $S_1$
and the properties of the maps.
A degree $1$ homological vector field $\hQ$ is defined as $\hQ = \sbv{S}{-}$.
The classical master equation shows that
$\hQ$ is a coboundary operator, $\hQ^2=0$.
%

From the above construction, we can prove that 
the mapping space $\Map(\calX, \calM)$ is a QP-manifold.
This structure is called an AKSZ sigma model.
If $\calX = T[1]X$, where $X$ is a manifold in $n+1$ dimensions,
the QP-structure on $\Map(\calX, \calM)$
is of degree $-1$.
In this case, a QP-structure on $\Map(T[1]X,  \calM)$
is equivalent to the Batalin-Vilkovisky formalism
of a topological sigma model including all ghosts and antifields.
$\sbv{-}{-}$ is the BV antibracket, 
and it is denoted by $\mbv{-}{-}$.

If $X$ is a manifold in $n$ dimensions,
the QP-structure on $\Map(\calX, \calM)$
is of degree $0$.
In this case, a QP-structure on $\Map(T[1]X,  \calM)$
is equivalent to the Hamiltonian BFV formalism
and $\sbv{-}{-}$ is an ordinary Poisson bracket.
Then we denote $\sbv{-}{-}$ by $\sbv{-}{-}_{PB}$.

\subsection{AKSZ sigma models with boundary}
\noindent
Recall the definition of a canonical transformation in subsection \ref{fluxduality}.
Let $\calM$ be a QP-manifold of degree $n$.
For any function $\alpha \in C^{\infty}(\calM)$ of degree $n$,
$e^{\delta_\alpha}$ is called a canonical transformation,
since $\sbv{e^{\delta_\alpha} f}{e^{\delta_\alpha} g} 
= e^{\delta_\alpha} \sbv{f}{g}$, where $f, g \in C^{\infty}(\calM)$.

For our purpose, we choose a special kind of canonical transformation,
called \textsl{canonical function with respect to a Lagrangian submanifold
$\calL$}.
\begin{definition}\label{CanonicalFunction}
Let $(\calM, \omega, Q)$
be a QP-manifold of degree $n$.
A function $\alpha$ of degree $n$
is called canonical function with respect to $\calL$, if
$e^{\delta_\alpha} \Theta |_{\calL} =0$, where $\calL$ is 
a Lagrangian submanifold
of $(\calM,\omega)$ and $|_{\calL}$ is the restriction to $\calL$.
\end{definition}

This transformation changes the target QP-manifold to
$(\calM, \omega, \Theta_{\alpha})$ with
$\Theta_{\alpha} = e^{\delta_\alpha} \Theta$.
Since the P-structure does not change, 
the new Q-structure function $S^{\prime}$ 
in the AKSZ sigma model becomes
\begin{eqnarray}
S^{\prime} &=& S_0 + S_1^{\prime}
\nonumber \\
&=& \iota_{\hat{D}} \mu_* {\rm ev}^* \vartheta
+ \mu_* \ev^*
e^{\delta_\alpha} \Theta.
\label{akszwithboundary}
\end{eqnarray}
If $\partial X = \emptyset$, 
$S^{\prime}$ also satisfies
the classical master equation,
since
$\sbv{e^{\delta_\alpha} \Theta}{e^{\delta_\alpha} \Theta}
= e^{\delta_\alpha} \sbv{\Theta}{\Theta} =0$.
If $\partial X \neq \emptyset$,
the boundary conditions are deformed by $\alpha$,
so that $S^\prime$ satisfies the classical master equation.
%
In this case, using Stokes' theorem, a straightforward computation gives
\begin{eqnarray}
\sbv{S^{\prime}}{S^{\prime}}= 
\iota_{\hat{D}} \mu_{\partial \calX *} \
(i_{\partial} \times {\rm id})^* \ 
\ev^* \vartheta 
+ \mu_{\partial \calX *} \
(i_{\partial} \times {\rm id})^* \ 
\ev^* e^{\delta_\alpha} \Theta,
\label{boundaryCME}
\end{eqnarray}
where $i_{\partial}$ is the inclusion map
$i_{\partial}: \partial \calX \longrightarrow \calX$
and
$\mu_{\partial \calX *}$ is the boundary integration
on $\partial \calX$ by the pullback $\mu_*$ 
by the map $i_{\partial}$.
Since the classical master equation,
$\sbv{S^{\prime}}{S^{\prime}}= 0$, 
must be satisfied for consistency of the theory,
the right hand side of \eeqref{boundaryCME} must vanish.
The condition is expressed on $\calM$
as follows.
\cite{Hofman:2002rv,Ikeda:2013wh}
\begin{proposition}\label{boundaryQstr}
We assume $\partial \calX \neq \emptyset$.
Let $(\calM, \omega, \Theta)$ be a QP manifold of degree $n$
and 
$\calL$ a Lagrangian submanifold of $\calM$,
which is the zero locus of the canonical $1$-form $\vartheta$,
where $\omega = - \delta \vartheta$.

Let $\alpha \in C^{\infty}(\calM)$ of degree $n$
be a canonical function
with respect to $\calL$,
i.e., 
$
e^{\delta_\alpha} \Theta |_{\calL} =0$.
Then, the classical master equation, $\sbv{S^{\prime}}{S^{\prime}}=0$,
is satisfied in an AKSZ sigma model (\ref{akszwithboundary}).
\end{proposition}
From proposition \ref{boundaryQstr},
the mathematical structure of an
AKSZ sigma model with boundary is described by
a quintuple $(\calM, \omega, \Theta, \calL, \alpha)$.

A canonical transformation by a canonical function $\alpha$ 
can be interpreted as an introduction of a boundary term in the sigma model
action $S$.

Let $S^{\prime\prime} = e^{- \delta_{\hat{\alpha}}} S^{\prime}$,
where $\hat{\alpha} = \mu_* \ev^* \alpha$.
Then, $S^{\prime}$ and $S^{\prime\prime}$ have equivalent 
geometric structures,
\begin{eqnarray}
S^{\prime\prime} &=& e^{- \delta_{\hat{\alpha}}} S^{\prime}
\nonumber \\
&=& 
e^{- \delta_{\hat{\alpha}}} S_0
+ \mu_* \ev^*
e^{- \delta_{\alpha}} e^{\delta_\alpha} \Theta
\nonumber \\
&=& e^{- \delta_{\hat{\alpha}}} S_0
+ \mu_* \ev^* \Theta.
\label{ChangeS}
\end{eqnarray}
Let us consider the special case, where $\alpha$ satisfies
$\sbv{\alpha}{\alpha}=0$. 
Then, since $e^{- \delta_{\hat{\alpha}}} S_0
= S_0- \sbv{S_0}{\hat{\alpha}}$,
${\alpha}$ generates a boundary term
$S_{\partial \calX} =
- \mu_{{\partial \calX}*} (i_{\partial} \times {\rm id})^* \ev^* {\alpha}$,
\begin{eqnarray}
S^{\prime\prime}
&=&
S_0- \sbv{S_0}{\hat{\alpha}}
+ \mu_* \, \ev^* \Theta
\nonumber \\
&=&
S_0 
-
L_{\hat{D}} \, \mu_* \ev^* {\alpha}
+ 
\mu_* \, \ev^* \Theta
\nonumber \\
&=&
S -
\mu_{\partial \calX *} \, (i_{\partial} \times {\rm id})^*
\ev^* {\alpha}.
\label{AKSZbulk}
\end{eqnarray}
Therefore, twisting of $S$ by $\alpha$ introduces
a boundary term induced by transgression of $\alpha$,
$S_{\partial \calX} =
- \int_{\partial \calX} \mu \ (i_{\partial} \times {\rm id})^* \ev^* {\alpha}$.

\subsection{Contravariant Courant sigma models}
\label{contraCSM}
\noindent
We construct the AKSZ sigma model induced from the
Poisson Courant algebroid.

Let us take a $3$-dimensional manifold $X$ with boundary 
$\partial X$.
The worldvolume is a supermanifold $\calX = T[1]X$.
Let $(\sigma^{\mu}, \theta^{\mu})$ be local coordinates
of degree $(0,1)$ on $\calX$.
Elements of
$\Map(\calX, \calM)$ are superfields, 
which we denote by boldface letters. 
For example, $\be \in \Gamma(\calX, \bbx^* \calM)$
corresponds to a local coordinate $e$ on $\calM$,
where $\bx: \calX \rightarrow M$.

The AKSZ construction on 
$\Map(\calX, T^*[2]T^*[1]M)$
gives the bulk AKSZ sigma model.
We denote the P-structure by
\begin{eqnarray}
\bomega 
&=& \int_{\calX}
\mu
\
(
\delta \bbx^{i} \wedge \delta \bbxi_{i}
+ \delta \bq^i \wedge \delta \bp_{i}
).
\end{eqnarray}
If $\alpha =0$,
the Q-structure function has the following form:
\begin{eqnarray}
S&=& \int_{\calX} \!
\mu
\left(
 \bxi_{i} \bbd \bbx^{i}
- \bbp_{i} \bbd \bbq^{i}
+ \pi^{ij}(\bbx) \bxi_i \bbp_j
- \frac{1}{2} \frac{\partial \pi^{jk}}{\partial x^i}(\bbx) 
\bbq^i \bbp_j \bbp_k
+ \frac{1}{3!} R^{ijk}(\bbx) \bbp_i \bbp_j \bbp_k
\right).
\label{3DcontraCourantSigmaModel}
\end{eqnarray}
We call \eeqref{3DcontraCourantSigmaModel}
the \textsl{Poisson Courant sigma model} 
or the \textsl{contravariant Courant sigma model}.
We take the variation of $S$,
\begin{eqnarray}
\delta S 
&=&
\int_{\calX}
\mu
\ \left( \delta \bxi_{i} \bbd \bbx^{i}
+ \bxi_{i} \bbd \delta \bbx^{i}
- \delta \bbp_{i} \bbd \bbq^{i}
- \bbp_{i} \bbd \delta \bbq^{i}
\right.
\nonumber \\ &&
\left. +
\delta
\left(
\pi^{ij}(\bbx) \bxi_i \bp_j
- \frac{1}{2} \frac{\partial \pi^{jk}}{\partial x^i}(\bbx) 
\bbq^i \bbp_j \bbp_k
+ \frac{1}{3!} R^{ijk}(\bbx) \bbp_i \bbp_j \bbp_k
\right) 
\right).
\end{eqnarray}
The equations of motion for $\bxi$ and $\bbq$ are obtained by 
integration by parts.
Since
\begin{eqnarray}
\delta S|_{\partial \calX} =
\int_{\partial \calX}
\mu_{\partial \calX}
\ \left(
 \bxi_{i} \delta \bbx^{i} + \bbp_{i} \delta \bbq^{i}
\right)|_{\partial \calX},
\label{boundaryintegration}
\end{eqnarray}
the boundary terms,
$\bxi_{i} \delta \bbx^{i} + \bbp_{i} \delta \bbq^{i}$,
must vanish
in order to derive consistent equations of motion.
This equation determines the boundary conditions.
It is satisfied, if for the Liouville $1$-form 
it holds $\ev^* \vartheta = 0$
on the boundary of the membrane 
in the target space.
This means, that the Lagrangian submanifold 
$\calL$ is the zero locus of $\vartheta$.

On the other hand, the boundary condition
must be consistent with the classical master equation
on the mapping space,
$\mbv{S}{S}=0$.
Direct computation gives
\begin{eqnarray}
\mbv{S}{S} =
\int_{\partial \calX}
\mu_{\partial \calX}
\left.
\left(
 \bxi_{i} \bbd \bbx^{i} - \bbp_{i} \bbd \bbq^{i}
+ \pi^{ij}(\bbx) \bxi_i \bbp_j
- \frac{1}{2} \frac{\partial \pi^{jk}}{\partial x^i}(\bbx) 
\bbq^i \bbp_j \bbp_k
+ \frac{1}{3!} R^{ijk}(\bbx) \bbp_i \bbp_j \bbp_k
\right) \right|_{\partial \calX}.
\label{boundarymasterequation}
\end{eqnarray}
We can take boundary conditions
$\bxi_{i}|_{\partial \calX} = 0$ and $\bbp_{i}{}|_{\partial \calX} =0$,
such that \eqref{boundaryintegration} and
\eqref{boundarymasterequation} vanish.
Therefore,
in terms of the target space, the structure is
$\xi_{i} = p_{i} =0$.
This corresponds to the Dirac structure \cite{lwx} $T^*M$
of the Poisson Courant algebroid $TM \oplus T^*M$.


Next, we consider a more nontrivial case 
with boundary term,
modifying the Q-structure by a canonical function $\alpha$.
As an example, we take
$\alpha = -\frac{1}{2} B_{ij}(x) q^i q^j$,
constructed from a $2$-form $B = \frac{1}{2} B_{ij}(x) dx^i \wedge dx^j$.
The Q-structure changes to
\begin{eqnarray}
S^{\prime\prime}&=& \int_{\calX}
\mu
\ \left(
 \bxi_{i} \bbd \bbx^{i}
- \bbp_{i} \bbd \bbq^{i}
+ \pi^{ij}(\bbx) \bxi_i \bbp_j
- \frac{1}{2} \frac{\partial \pi^{jk}}{\partial x^i}(\bbx) 
\bbq^i \bbp_j \bbp_k
+ \frac{1}{3!} R^{ijk}(\bbx) \bbp_i \bbp_j \bbp_k
\right)
\nonumber \\
&& +
\int_{\partial \calX}
\mu_{\partial \calX}\
\frac{1}{2} B_{ij}(\bx) \bq^i \bq^j,
\label{3DcontraCourantSigmaModelboundary}
\end{eqnarray}
where 
we have used the expression 
$S^{\prime\prime} = e^{- \delta_{\hat{\alpha}}}S^{\prime}$
in \eeqref{AKSZbulk}.
%
The boundary term deforms the boundary conditions.
The variation $\delta S^{\prime\prime}$
restricted to the boundary is
\begin{eqnarray*}
\delta S^{\prime\prime}|_{\partial \calX}
=
\int_{\partial \calX}
\mu_{\partial \calX}
\ \left[\left( \bxi_{i}
+ \frac{1}{2} \frac{\partial B_{jk}(\bbx)}{\partial \bbx^i}
\bq^{j} \bq^{k} \right)
\delta \bbx^{i}
+ \left( \bbp_{i}
-
B_{ij}(\bbx) \bq^{j}
\right)
\delta \bbq^{i}
+ \cdots
\right].
\end{eqnarray*}
Since these terms must vanish, consistent boundary conditions are as follows:
\begin{eqnarray}
\bxi_{i}|_{\partial \calX} =
- 
\frac{1}{2} \frac{\partial B_{jk}(\bbx)}{\partial \bbx^i}
\bq^{j} \bq^{k}|_{\partial \calX},
\qquad
\bp_i|_{\partial \calX} &=& B_{ij}(\bbx) \bq^{j} |_{\partial \calX}.
\label{WZPoissonboundary}
\end{eqnarray}
The master equation, $\mbv{S^{\prime\prime}}{S^{\prime\prime}}=0$, requires
another consistency condition,
i.e., the integrand of $S_1$ is zero on the boundary,
\begin{eqnarray}
\left.
\left(
\pi^{ij}(\bbx) \bxi_i \bbp_j
- \frac{1}{2} \frac{\partial \pi^{jk}}{\partial x^i}(\bbx) 
\bbq^i \bbp_j \bbp_k
+ \frac{1}{3!} R^{ijk}(\bbx) \bbp_i \bbp_j \bbp_k
\right) \right|_{\partial \calX}
= 0.
\label{boundaryS1}
\end{eqnarray}
Similarly, (\ref{WZPoissonboundary}) and (\ref{boundaryS1}) can 
be converted to a condition on the target space, such that
\begin{eqnarray}
\Theta = \pi^{ij}(x) \xi_i p_j
- \frac{1}{2} \frac{\partial \pi^{jk}}{\partial x^i}(x) 
q^i p_j p_k
+ \frac{1}{3!} R^{ijk}(x) p_i p_j p_k
= 0
\label{WZPS1}
\end{eqnarray}
holds on the Lagrangian submanifold $\calL^{\prime}
$, defined by
\begin{eqnarray}
\xi_{i}=
- 
\frac{1}{2} \frac{\partial B_{jk}(x)}{\partial x^i}
q^{j} q^{k},
\qquad
p_i &=& B_{ij}(x) q^{j}.
\label{WZPLagrangian}
\end{eqnarray}
%
Substituting (\ref{WZPLagrangian}) into
(\ref{WZPS1}), we obtain the following geometric structure on
$\calL^{\prime}$:
\begin{eqnarray}
&& 
\pi^{ij}(x) \xi_i p_j
- \frac{1}{2} \frac{\partial \pi^{jk}}{\partial x^i}(x) 
q^i p_j p_k
+ \frac{1}{3!} R^{ijk}(x) p_i p_j p_k
\nonumber \\
&& =
\left(
- \frac{1}{2} \pi^{lm}
\frac{\partial B_{ij}}{\partial x^l}
B_{mk}
- \frac{1}{2} 
\frac{\partial \pi^{lm}}{\partial x^i}
B_{jl}
B_{km}
- \frac{1}{3!}
R^{lmn} B_{il}
B_{jm} B_{kn}
\right)
q^{i} q^{j} q^{k} =0.
\label{twistedPoisson}
\end{eqnarray}
\eeqref{twistedPoisson} is equivalent to
\begin{eqnarray}
[B,B]_{\pi} = \wedge^3 B^{\flat} R.
\label{twisted2formbyR}
\end{eqnarray}
\footnote{
The Koszul bracket $[-,-]_{\pi}$ is extended to 
the space of all differential forms
as the Lie bracket satisfying the Leibniz rule.
The homomorphism $B^{\flat}: TM \rightarrow T^*M$
is defined by $B^{\flat}(X) = B_{ij} X^i(x) \partial/\partial x^j$,
where $X = X^i(x) \partial/\partial x^i$
is a vector field.
}
The commutator of a $2$-form $B$ with respect to the Koszul bracket 
is twisted by a $3$-vector field $R$.
If $B = \pi^{-1}$, 
\eeqref{twistedPoisson} becomes
\begin{eqnarray}
H = dB = \wedge^3 B^{\flat} R.
\end{eqnarray}

Next, we construct the boundary action on $T[1] \partial X$
by integrating out the superfield $\bxi_i$
and using the Stokes' theorem.
Suppose $\pi$ is nondegenerate, i.e., 
$\pi^{-1}$ is symplectic. 
Integrating out $\bxi_i$ from the action
(\ref{3DcontraCourantSigmaModelboundary}), 
we obtain the equations of motion 
$\bbp_i = -\pi^{-1}_{ij} \bbd \bbx^j$.
By substituting this equation to 
\eeqref{3DcontraCourantSigmaModelboundary}, the action becomes
the boundary (twisted) AKSZ sigma model with WZ term in two dimensions,
\begin{eqnarray}
S &=& \int_{\partial \calX}
\mu_{\partial \calX} \ 
(\pi^{-1})_{ij} \bbq^i \bbd \bbx^{j}
+ \frac{1}{2} B_{ij}(\bx) \bq^i \bq^j
\nonumber \\ &&
- \int_{\calX}
\mu \ 
\frac{1}{3!} R^{ijk}(\bbx) 
(\pi^{-1})_{il}
(\pi^{-1})_{jm}
(\pi^{-1})_{kn} \bbd \bbx^{l} \bbd \bbx^{m} \bbd \bbx^{n}.
\label{2Dcontraactionboundary}
\end{eqnarray}
This is the Poisson sigma model 
\cite{Ikeda:1993aj,Ikeda:1993fh,Schaller:1994es}
deformed by a WZ term \cite{Klimcik:2001vg}.

The action without ghosts can be obtained as follows.
We expand the superfields in components as 
\begin{eqnarray}
\bPhi(\sigma, \theta) = \Phi^{(0)}(\sigma)
+ \Phi^{(1)}_{\mu}(\sigma) \theta^{\mu} 
+ \frac{1}{2} \Phi^{(2)}_{\mu\nu}(\sigma) \theta^{\mu} \theta^{\nu}.
\end{eqnarray}
If we integrate over $\theta^{\mu}$ and drop 
ghost fields with nonzero degrees, we get the physical action:
\begin{eqnarray}
S &=& - \int_{\partial X}\left(
(\pi^{-1})_{ij} q^{i} \wedge d x^{j}
- \frac{1}{2} B_{ij}(x) q^{i} \wedge q^{j} \right)
\nonumber \\ &&
- \int_{X}
\frac{1}{3!} R^{ijk}(x) 
(\pi^{-1})_{il}
(\pi^{-1})_{jm}
(\pi^{-1})_{kn} d x^{l} \wedge d x^{m} \wedge d x^{n},
\label{2Dcontraactionboundary}
\end{eqnarray}
where $x^i = x^{(0)i}$ and $q^i = d \sigma ^{\mu} q_{\mu}^{(1)i}$.
If we add the kinetic term, we obtain 
a string sigma model action with R-flux,
\begin{eqnarray}
S &=& \frac{1}{2} \int_{\partial X}
G_{ij}(x)
d x^{i} \wedge * d x^{j}
- \int_{\partial X}
\left(
(\pi^{-1})_{ij} q^{i} \wedge d x^{j}
- \frac{1}{2} B_{ij}(x) q^{i} \wedge q^{j}
\right)
\nonumber \\ &&
- \int_{X}
\frac{1}{3!} R^{ijk}(x) 
(\pi^{-1})_{il}
(\pi^{-1})_{jm}
(\pi^{-1})_{kn} d x^{l} \wedge d x^{m} \wedge d x^{n}.
\label{2Dcontraactionboundary}
\end{eqnarray}

\subsection{Duality of Courant sigma models}
\label{CSMduality}
\noindent
We have discussed duality transformations of the standard and the Poisson 
Courant algebroids in section 4.
In this subsection, 
we derive the same result from the analysis of the corresponding sigma models.

We perform the duality transformation on the level of sigma models.
The AKSZ construction on a three-dimensional manifold $X$ with boundary
gives rise to two Courant sigma models,
one with the Poisson Courant algebroid structure with $R$-flux,
constructed in the previous subsection,
and one
with the standard Courant algebroid structure with $H$-flux.

The BV action of the Poisson Courant sigma model 
is \eeqref{3DcontraCourantSigmaModel}.
On the other hand, from the AKSZ construction, 
the BV action of the standard Courant sigma model is 
\begin{eqnarray}
S &=& \int_{\calX}
\mu
\ \left(
 \bxi_{i} \bbd \bbx^{i}
- \bbp_{i} \bbd \bbq^{i}
+ \bxi_i \bbq^i
+ \frac{1}{3!} H_{ijk}(\bbx) \bbq^i \bbq^j \bbq^k
\right).
\label{3DstandardCourantSigmaModel}
\end{eqnarray}
{
We consider the following twisting of the standard Courant sigma model 
by applying the twist \eeqref{twistedHRfluxTheta} as
\begin{eqnarray}
S^{\prime} &=& \int_{\calX}
\mu
\ \left(
 \bxi_{i} \bbd \bbx^{i}
- \bbq^{i} \bbd \bbp_{i}
\right)
+ 
e^{\delta_{\alpha_p}} e^{-\delta_{\alpha_q}} e^{\delta_{\alpha_p}}
\int_{\calX}
\mu
\left( \bxi_i \bbq^i
+ \frac{1}{3!} H_{ijk}(\bbx) \bbq^i \bbq^j \bbq^k
\right).
\end{eqnarray}
This is equivalent to
\begin{align}
S^{\prime\prime} =
e^{-\delta_{\alpha_p}} e^{\delta_{\alpha_q}}
e^{-\delta_{\alpha_p}} 
\int_{\calX}
\mu
\ \left(
 \bxi_{i} \bbd \bbx^{i}
- \bbq^{i} \bbd \bbp_{i}
\right)
+ 
\int_{\calX}
\mu
\left( \bxi_i \bbq^i
+ \frac{1}{3!} H_{ijk}(\bbx) \bbq^i \bbq^j \bbq^k
\right),
\end{align}
where $\alpha_p$ and $\alpha_q$ are understood as
$\mu_{\partial \calX *} \ev^* \alpha_p$
and $\mu_{\partial \calX *} \ev^* \alpha_q$.
We have
\begin{eqnarray}
e^{-\delta_{\alpha_p}} 
e^{\delta_{\alpha_q}}
e^{-\delta_{\alpha_p}}
S_0 
&=& S_0 
+ \int_{\partial \calX}
\mu_{\partial \calX}\
\left(
\frac{1}{2} \pi^{ij}(\bx) \bp_i \bp_j
+ \bp_i \bq^i
\right)
\nonumber \\
&& 
- 
\int_{\partial \calX}
\mu_{\partial \calX}\
\left(
- \frac{1}{2} \pi^{ij}(\bx) \bp_i \bp_j 
+ \pi^{ij}(\bx) \bp_i \bp_j 
\right)
\nonumber \\
&=& S_0 
+ \int_{\partial \calX}
\mu_{\partial \calX}\
\left(
\bp_i \bq^i
\right).
\end{eqnarray}
}
Therefore, by this twist,
the action becomes the Courant sigma model with boundary term,
\begin{eqnarray}
S^{\prime\prime} &=& 
\int_{\calX}
\mu
\ \left(
 \bxi_{i} \bbd \bbx^{i}
- \bbq^{i} \bbd \bbp_{i}
+ \bxi_i \bbq^i
+ \frac{1}{3!} H_{ijk}(\bbx) \bbq^i \bbq^j \bbq^k
\right)
+ \int_{\partial \calX}
\mu_{\partial \calX}\
\left(
\bp_i \bq^i
\right).
\end{eqnarray}
{
From \eeqref{HtoRcanonical1}--\eeqref{HtoRcanonical4}, 
redefining the superfields as
\begin{eqnarray}
&& \bx^i = \bxpr^{i},
\\
&& \bq^i = \pi^{ij}(\bx) \bppr_j,
\\
&& \bp_i = 
- \pi^{-1}_{ij}(\bx) \bqpr^j,
\\
&& \bxi_i = \bxipr_i 
+ \frac{\partial \pi^{jk}}{\partial x^i} \pi^{-1}_{kl}(\bx) \bppr_j \bqpr^l,
\end{eqnarray}
we can simplify the total action as
\begin{eqnarray}
S^{\prime\prime} &=& 
\int_{\calX}
\mu
\ \left(
 \bxipr_{i} \bbd \bbxpr^{i}
- \bbppr_{i} \bbd \bbqpr^{i}
\right.
\nonumber \\ 
&&
\left.
+ \pi^{ij}(\bbx) \bxipr_i \bbppr_j
- \frac{1}{2} \frac{\partial \pi^{jk}}{\partial x^i}(\bbx) 
\bbqpr^i \bbppr_j \bbppr_k
+ \frac{1}{3!} R^{ijk}(\bbx) \bbppr_i \bbppr_j \bbppr_k
\right).
\end{eqnarray}
The resulting action is 
the same as the BV action of the Poisson 
(contravariant) Courant sigma model
\eeqref{3DcontraCourantSigmaModel},
and we obtain the relation between $H$ and $R$, 
$R = \wedge^3 \pi^{\sharp} H$, again.
Therefore, in the theory of the topological membrane,
the duality transformation between $H$-flux and $R$-flux is a
change of boundary conditions.
}

\section{Current algebras}\label{contracurrentalgebra}
\label{sectioncurrentalgebra}
\noindent
In this section, we consider 
a current algebra \`a la Alekseev and Strobl 
\cite{Alekseev:2004np,Bonelli:2005ti,Ikeda:2011ax,Ikeda:2013vga}
corresponding to the Poisson Courant algebroid
in two-dimensional spacetime.

\subsection{Poisson brackets with fluxes from target QP-structures}
\noindent
In this subsection, we briefly review the Hamiltonian method to construct
a Poisson bracket of canonical variables with fluxes from general target 
QP-structures \cite{Ikeda:2013vga}.

Let $(\calM, \omega)$ be a P-manifold of degree $n-1$.
We take a worldvolume $X = \Sigma \times \bR$ in $n$ dimensions and
a \textsl{space} supermanifold $\calX = T[1]\Sigma$, since we consider
the Hamiltonian formalism.

The simplest method to determine a Poisson bracket is 
to construct it from $\bomega = \mu_* \ev^* \omega$.
Since $\Sigma$ is in $n-1$ dimensions and $\omega$ is of degree $n-1$, 
the graded symplectic structure 
$\bomega$ is of degree zero
due to the integration $\mu_*$.
Then, we obtain a graded Poisson bracket of degree zero,
that is, an ordinary Poisson bracket $\sbv{-}{-}_{PB}$.
However, the Poisson brackets 
obtained in this way cannot include
fluxes,
since the target space geometric datum $\Theta$ is not used.
In \cite{Alekseev:2004np}, 
the symplectic form was deformed 
by $b$-transformation to include H-flux.

Here, we use another method to incorporate the geometric datum
$\Theta$.
First we consider a QP-manifold of degree $n$, $(\calM, \omega, \Theta)$,
as an extended target space.
A Lagrangian submanifold $\calL$ with respect to $\omega$ 
is regarded as the target space of physical canonical quantities.
$\omega$ defines a graded Poisson bracket, $\sbv{-}{-}$, of degree $-n$.

Then, we consider the derived bracket
$\sbv{\sbv{-}{\Theta}}{-}$, which is of degree $-n+1$,
and use the following fact.
\begin{theorem}
Let $(\calM, \omega, Q)$ be a QP-manifold, 
$\calL$ a Lagrangian submanifold with respect to $\omega$,
and $\proj: \calM \rightarrow \calL$ the natural projection.
If the derived bracket is restricted to $\calL$, then it 
gives the graded Poisson bracket 
$\sbv{f}{g}_{\calL} \equiv \sbv{\sbv{\proj^* f}{\Theta}}{\proj^* g}\Big|_{\calL}
$
for functions $f$ and $g$ on $\calL$.
\end{theorem}
We can easily prove that 
$\sbv{f}{g}_{\calL}$ is antisymmetric
and satisfies both the Leibniz rule and the Jacobi identity
using $\sbv{\proj^* f}{\proj^* g} =0$.

In order to define a Poisson bracket on the mapping space, we transgress
the derived bracket to the mapping space $\Map(\calX, \calM)$.
Then, the derived bracket $\sbv{\sbv{-}{S_1}}{-}$ of degree zero, 
with $S_1 = \mu_* \ev^* \Theta$,
defines the Poisson bracket on the mapping space,
\begin{eqnarray}
\sbv{F}{G}_{PB} = \sbv{\sbv{F}{S_1}}{G}\Big|_{\hatcalL},
\end{eqnarray}
where $F, G$ are functions on the mapping space.
Precisely speaking, 
for a function $f \in C^{\infty}(\calL)$,
we introduce test functions
$\epsilon
$ on $\calX$ of degree $n-1 - |f|$.
A Lagrangian submanifold 
of $\Map(\calX, \calM)$ is denoted by
$\hatcalL$, and the projection map is
$\hatpi: \Map(\calX, \calM)\rightarrow \hatcalL$.
We can prove that 
the restriction of the derived bracket to a Lagrangian submanifold
$\hatcalL$,
%
$
\hatpi_* \sbv{\sbv{-}{S_1}}{-}
= \sbv{\sbv{-}{S_1}}{-}|_{\hatcalL}$,
becomes an ordinary Poisson bracket 
$%
\sbv{-}{-}_{PB}$,
\begin{eqnarray}
\sbv{\mu_* \epsilon_{1} \ev^* \proj^* f}{
\mu_* \epsilon_{2} \ev^* \proj^* g}_{PB}
&=& \hatpi_* \sbv{\sbv{\mu_* \epsilon_{1} \ev^* \proj^* f}{S_1}}{
\mu_* \epsilon_{2} \ev^* \proj^* g}.
\end{eqnarray}
The Poisson bracket depends not only on $\Theta$, but also on the choice 
of the Lagrangian submanifold $\hatcalL$.

Note that we can use the following formula to connect 
the target space computations to the superfield computations,
\begin{eqnarray}
\sbv{\sbv{\mu_* \epsilon_{1} \ev^* \proj^* f}{S_1}}{
\mu_* \epsilon_{2} \ev^* \proj^* g}
&=& 
\mu_* \epsilon_{1} \epsilon_{2} \ev^* 
\sbv{\sbv{\proj^* f}{\Theta}}{\proj^* g},
\label{transgressionPoissonbracket1}
\end{eqnarray}
where $f, g \in C^{\infty}(\calL)$.

Simple candidates for $\hatcalL$ are canonical Lagrangian submanifolds,
which we denote by $\hatcalL_0$.
For instance, in the case of the Courant algebroids, 
two simple Lagrangian submanifolds in $\Map(\calX, T^*[2]T^*[1]M)$
are $\Map(\calX, T^*[1]M) 
= \{(\bx^i, \bp_i, \bq^i, \bxi_i) | \bxi_i = \bq^i =0 \}$
or $\Map(\calX, T[1]M) = \{(\bx^i, \bp_i, \bq^i, \bxi_i) 
| \bxi_i = \bp_i =0 \}$.
Generally,  
we cannot obtain the twisted Poisson bracket with fluxes by 
simple restriction to these canonical Lagrangian submanifolds.
We do a special twisting of functions on the mapping space,
before restricting the space to a canonical Lagrangian submanifold,
where the twisting does not depend on fluxes.
This procedure derives a Poisson bracket with flux.

The space of functions on a canonical Lagrangian submanifold is 
constructed as\\
$C^{\infty}(\hatcalL_0) \equiv
\{\hatpi_{*} \mu_* \epsilon \ \ev^* \proj^* f | f \in C^{\infty}(\calL_0) \}$.
We twist the functions to be 
$e^{\delta_{\alpha}} \mu_* \epsilon \ \ev^* \proj^* f
= e^{\delta_{\alpha}} \int_{\calX} \mu \ \epsilon(\sigma, \theta) 
f(\bp(\sigma, \theta))$.
The twisted derived bracket
$\sbv{\sbv{e^{\delta_{\alpha}} \mu_* \epsilon \ \ev^* \proj^* f}
{S_1}}{e^{\delta_{\alpha}} \mu_* \epsilon \ \ev^* \proj^* g}$
restricted to $\hatcalL_0$
gives rise to a Poisson bracket with fluxes,
if we choose $\alpha$ properly.
%
%
%
%
%
The formula 
is 
\begin{eqnarray}
\sbv{e^{\delta_{\alpha}} \mu_* \epsilon_{1} \ev^* \proj^* f}
{e^{\delta_{\alpha}} \mu_* \epsilon_{2} \ev^* \proj^* g}_{PB}
&\equiv & 
\hatpi_{*} \sbv{\sbv{e^{\delta_{\alpha}} \mu_* \epsilon_{1} \ev^* \proj^* f}
{S_1}}{e^{\delta_{\alpha}} \mu_* \epsilon_{2} \ev^* \proj^* g}.~~~~~~~
\label{transgressionPoissonbracket1}
\end{eqnarray}


Assume that a nontwisted Poisson bracket
$\sbv{-}{-}_{PB}$ on the canonical Lagrangian submanifold
is nondegenerate, that is, symplectic\footnote{This condition is satisfied, if $\calM$ is a double 
graded cotangent bundle of a graded manifold $\calN$, for example, 
$\calM = T^*[n]T^*[n-1]\calN$. All QP-manifolds in this paper are 
double cotangent bundles.}.
Then, there exists a graded symplectic form 
$\bomega_{\hatcalL_0} = \mu_* \ev^* \omega_{\calL_0}$
on $\hatcalL_0$ 
corresponding to this Poisson bracket.
We choose $\alpha_0 = \iota_{\hat{D}} \mu_* \ev^* \vartheta_{\calL_0}$
as the twisting function,
where $\vartheta_{\calL_0}$ is the Liouville $1$-form,
such that $\omega_{\calL_0} = - \delta \vartheta_{\calL_0}$.
Note that $\alpha_0$ does not depend on the fluxes.

%
Here, 
we have twisted the space of functions on 
the mapping space and restricted it to 
the canonical Lagrangian submanifold.
We could also obtain the same Poisson bracket
by twisting the canonical Lagrangian submanifold
with a canonical transformation $e^{-\delta_{\alpha}}$
and restricting the derived bracket to the twisted Lagrangian submanifold.

We demonstrate the procedure in the $H$-flux case.

\begin{example}[Poisson brackets twisted by H-flux]
\label{PBofHflux}
\noindent
We construct the Poisson bracket of the standard Courant algebroid 
with $H$-flux on the cotangent space of the loop space $\Map(S^1,T^*M)$.
First we determine the Poisson brackets of the canonical quantities.

Let us consider the QP-manifold of the standard Courant algebroid,
$\calM = T^*[2]T^*[1]M$,
in section \ref{CAandsupergeometry}.
Take local Darboux coordinates $(x^i, p_i, q^i, \xi_i)$ 
of degree $(0,1,1,2)$.
The canonical graded symplectic structure is expressed by
$\omega = \delta x^i \wedge \delta \xi_i + \delta p_i \wedge \delta q^i$.
We take a canonical Lagrangian submanifold 
with respect to $\omega$,
$\calL_0 = \{(x^i, p_i) | \xi_i = q^i =0 \}$.

Since the Q-structure function of the standard Courant algebroid is
$\Theta = \xi_i q^i + \frac{1}{3!} H_{ijk}(x) q^i q^j q^k$,
the derived brackets for the canonical quantities on $\calL_0$
are
\begin{eqnarray}
\sbv{\sbv{x^i}{\Theta}}{x^j} &=& 0,
\label{standtargetcommu1}
\\
\sbv{\sbv{x^i}{\Theta}}{p_j} 
&=& 
\delta^i{}_j,
\label{standtargetcommu2}
\\ 
\sbv{\sbv{p_i}{\Theta}}{p_j}
&=& 
- H_{ijk}(x) q^k.
\label{standtargetcommu3}
\end{eqnarray}
Next, we consider the mapping space. 
Take the supermanifold $\calX = T[1]S^1$ with
local coordinates $(\sigma, \theta)$.
Local coordinates on $\Map(T[1]S^1, T^*[2]T[1]M)$
are superfields $\bx^i(\sigma, \theta): T[1]S^1 \rightarrow M$
and $\bq^i(\sigma, \theta) \in \Gamma(T^*[1]S^1 \otimes \bx^* (T_x[1]M))$ 
of degree $(0, 1)$
and canonical conjugates
$\bxi_i(\sigma, \theta)$
and $\bp_i(\sigma, \theta)$
of degree $(2, 1)$,
$\bxi_i(\sigma, \theta) \in \Gamma(T^*[1]S^1 \otimes \bx^*(T^*_x[2]M))$
and 
$\bp_i(\sigma, \theta) \in \Gamma(T^*[1]S^1 \otimes \bx^*(T^*_q[2]T_x[1]M))$.

The transgression 
of \eeqref{standtargetcommu1}--\eqref{standtargetcommu3} induces
the derived bracket on superfields.
The concrete expression is
\begin{eqnarray}
\sbv{\sbv{\bx^i(\sigma, \theta)}{S_1}}
{\bx^j(\sigma^{\prime}, \theta^{\prime})} &=& 0,
\label{standtargetcommu11}
\\
\sbv{\sbv{\bx^i(\sigma, \theta)}{S_1}}
{\bp_j(\sigma^{\prime}, \theta^{\prime})} 
&=& 
- \delta^i{}_j
\delta(\sigma-\sigma^\prime)
\delta(\theta-\theta^\prime),
\label{standtargetcommu12}
\\ 
\sbv{\sbv{\bp_i(\sigma, \theta)}{S_1}}
{\bp_j(\sigma^{\prime}, \theta^{\prime})}
&=& 
H_{ijk}(\bx) \bq^k(\sigma, \theta)
\delta(\sigma-\sigma^\prime)
\delta(\theta-\theta^\prime).
\label{standtargetcommu13}
\end{eqnarray}
The Liouville $1$-form on the Lagrangian submanifold is
$\alpha_0 = \iota_{\hat{D}} \mu_* \ev^* \vartheta_{\calL}
= - \int \mu \, \bp_i \bbd \bx^i$.
Twisting by the Liouville $1$-form 
$\alpha_0$ gives rise to the transformation
$\bq^k \rightarrow \bq^k - \bbd \bx^k$.
If we reduce to the canonical Lagrangian submanifold
$\hatcalL_{0}$ defined by $\bxi_i = \bq^i =0$,
we obtain 
\begin{eqnarray}
\{\bx^i(\sigma, \theta),\bx^j(\sigma^{\prime}, \theta^{\prime})\}_{PB} 
&=& 0,
\label{superfieldcommuH1}
\\
\{\bx^i(\sigma, \theta) ,\bp_j(\sigma^{\prime}, \theta^{\prime})\}_{PB} 
&=& 
- \delta^i{}_j
\delta(\sigma-\sigma^\prime)
\delta(\theta-\theta^\prime), 
\label{superfieldcommuH2}
\\ 
\{\bp_i(\sigma, \theta), \bp_j(\sigma^{\prime}, \theta^{\prime})\}_{PB}
&=& 
- H_{ijk}(\bx(\sigma, \theta)) \bbd \bx^k(\sigma, \theta)
\delta(\sigma-\sigma^\prime)
\delta(\theta-\theta^\prime).
\label{superfieldcommuH3}
\end{eqnarray}
We expand the superfields by the local coordinate $\theta$ on $T[1]S^1$,
\begin{eqnarray}
\bPhi(\sigma, \theta)= \Phi^{(0)}(\sigma) + \Phi^{(1)}(\sigma) \theta.
\end{eqnarray}
The degree zero component 
in the expansion
is the physical field
(and degree nonzero components are ghost fields).
In this example, physical fields are $x^i(\sigma) = x^{(0)i}(\sigma)$
and $p_i(\sigma) = p^{(1)}_i(\sigma)$.
The Poisson brackets of the physical canonical quantities 
are degree zero components of \eeqref{superfieldcommuH1}--\eqref{superfieldcommuH3}:
\begin{eqnarray}
\{x^i(\sigma),x^j(\sigma^{\prime})\}_{PB} 
&=& 0,
\label{2dcanoicalcommutation11}
\\
\{x^i(\sigma) ,p_j(\sigma^{\prime})\}_{PB} 
&=& 
\delta^i{}_j
\delta(\sigma-\sigma^\prime),
\label{2dcanoicalcommutation12}
\\ 
\{p_i(\sigma), p_j(\sigma^{\prime})\}_{PB}
&=& 
- H_{ijk}(x) \partial_{\sigma} x^k
(\sigma)
\delta(\sigma-\sigma^\prime).
\label{2dcanoicalcommutation13}
\end{eqnarray}
These are the Poisson brackets of the canonical quantities with H-flux
in \cite{Alekseev:2004np}.
The symplectic form 
of Alekseev-Strobl type,
which induces
(\ref{2dcanoicalcommutation11})--(\ref{2dcanoicalcommutation13}),
is
\beq
    \omega=
\int_{S^1}d\sigma \ \delta x^i \wedge \delta p_i
+
\frac{1}{2} \int_{S^1} d\sigma 
H_{ijk}(x) \partial_{\sigma} x^i
\delta x^j \wedge \delta x^k.
\label{2dstandsymplectic}
\eeq

\end{example}

\subsection{Current algebras from target QP-Structures}
\noindent
A current algebra is a Poisson algebra on the mapping space.
Consider a subspace of $C^{\infty}(\calM)$, which is 
closed under the derived bracket
$\sbv{\sbv{-}{\Theta}}{-}$. 
This subspace becomes a Poisson algebra
on the mapping space 
after transgression.
The finite degree subspace,
$\sum_{i=0}^{n-1} C_{i}(\calM)
= \{f \in C^{\infty}(\calM)| |f| \leq n-1 \}$,
is closed under the derived bracket.
%
If 
$C_{n-1}(\calM)$ is transgressed to
$\Map(\calX, \calM)$, twisted by $\alpha_0$ and 
restricted to the Lagrangian submanifold
$\hatcalL_{0}$, we obtain a current algebra with fluxes.

For the $H$-flux case, we demonstrate the construction in the following
example.

\begin{example}[Alekseev-Strobl current algebras]\label{AScurrentalgebra}

Based on the result in example \ref{PBofHflux},
we take
$C_0 \oplus C_1
= \{f \in C^{\infty}(T^*[2]T^*[1]M)| |f| \leq 1 \}$
as space of functions.
Elements of $C_0 \oplus C_1$ are
functions of degree zero, $j_{0(f)} = f(x)$
and functions of degree one,
$j_{1(X + \alpha)} = \alpha_i(x) q^i + X^i(x) p_i$.
$j_{1(X + \alpha)}$ is equivalent to a section of $TM \oplus T^*M$,
since it corresponds to
$\alpha_i(x) d x^i + X^i(x) \frac{\partial}{\partial x^i}$.
Let
$j_{(0)(g)} = g(x)$ and
$j_{(1)(Y + \beta)} = \beta_i(x) q^i + Y^i(x) p_i$.
The derived brackets of functions are easily computed,
\begin{eqnarray}
\sbv{\sbv{j_{0(f)}}{\Theta}}{j_{0(g)}} &=& 0,
\nom\\
\sbv{\sbv{j_{1(X+ \alpha)}}{\Theta}}{j_{0(g)}}
 &=& -X^{i}\frac{\del j_{0(g)}}{\del x^i}
 = - \rho(X + \alpha) j_{0(g)},
\nom\\
\sbv{\sbv{j_{1(X+\alpha)}}{\Theta}}{j_{1(Y+ \beta)}}
    &=& - j_{1 (\couranth{X + \alpha}{Y + \beta})},
\label{bigsupercommutation}
\end{eqnarray}
where $\couranth{X+\alpha}{Y+\beta}$ is the Dorfman bracket 
on the standard Courant algebroid with flux $H$,
\begin{eqnarray}
\couranth{X + \alpha}{Y + \beta}
&=& [X, Y] + L_X \beta - \iota_Y d \alpha + \iota_X \iota_Y H.
\label{2dsupercommutation}
\end{eqnarray}
for $X, Y \in \Gamma(TM)$, $\alpha, \beta \in \Gamma(T^*M)$.
%

Currents are identified with twisted functions on the Lagrangian submanifold of 
the mapping space.
In order to construct currents, 
we apply the transgression map to $j_0$ and $j_1$. Then,
we twist them
by $\alpha_0$ and finally restrict the resulting functions 
to the Lagrangian submanifold defined by $\bxi_i = \bq^i =0$.
The corresponding currents are 
\begin{eqnarray}
&& \bJ_{(0)(f)}(\epsilon_{(1)})
= \tilpi_* e^{\delta_{\alpha_0}} \mu_* \epsilon_{(1)} \ev^*  j_{(0)(f)}
= \int_{T[1]S^1} \mu \epsilon_{(1)} f(\bx),
\nonumber \\
&&
\bJ_{(1)(X+ \alpha)}(\epsilon_{(0)})
= \tilpi_* e^{\delta_{\alpha_0}} \mu_* \epsilon_{(0)} \ev^* j_{(1)(u,\alpha)}
= \int_{T[1]S^1} \mu \epsilon_{(0)} 
(- \alpha_i(\bx) \bbd \bx^i + X^i(\bx) \bp_i),
\end{eqnarray}
where $\epsilon_{(i)} = \epsilon_{(i)}(\sigma, \theta)$ 
is a test function  of degree $i$ on the super circle $\calX = T[1]S^1$.
The integrands of degree zero components of $\bJ_0$ and $\bJ_1$
are
\begin{eqnarray}
&& J_{(0)(f)}(\sigma) = f(x(\sigma)),
\qquad
J_{(1)(X+ \alpha)}(\sigma)
= \alpha_i(x) \partial_{\sigma} x^i(\sigma) + X^i(x) p_i(\sigma),
\end{eqnarray}
which are the correct AS currents.

We compute the Poisson algebra of these supergeometric currents
from the Poisson brackets of canonical quantities $(\bx^i, \bp_i)$
obtained in example \ref{PBofHflux}:
\beqa
    \{\bJ_{0(f)}(\epsilon),
\bJ_{0(g)}(\epsilon^{\prime})
\}_{PB}
&=&
0,
\label{2dstandalekseevstrtoblsupercommutation1} 
\\
     \{\bJ_{1(X+\alpha)}(\epsilon),
\bJ_{0(g)}(\epsilon^{\prime})\}_{PB}
&=&
\rho(X + \alpha) \bJ_{0(g)}
(\epsilon \epsilon^{\prime}),
\label{2dstandalekseevstrtoblsupercommutation2}
\\
    \{\bJ_{1(X+\alpha)}(\epsilon),
\bJ_{1(Y+\beta)}(\epsilon^{\prime})\}_{PB}
    &=&
\bJ_{1 (\couranth{X + \alpha}{Y + \beta})}(\epsilon \epsilon^{\prime})
\nonumber \\
&&
+ \int_{T[1]S^1} \mu \bbd \epsilon_{(0)} \epsilon^{\prime}_{(0)}
\bracket{X + \alpha}{Y + \beta}(\bx),
\label{2dstandalekseevstrtoblsupercommutation3}
\eeqa
where
$\bJ_{0(g)}^{\prime} = \int_{T[1]S^1} \mu \epsilon_{(1)} g(\bx)$,
$\bJ_{1(Y+\beta)}^{\prime} =
\int_{T[1]S^1} \mu \epsilon_{(0)} 
(- \beta_i(\bx) \bbd \bx^i + Y^i(\bx) \bp_i)$.
The AS current algebra is given by 
the physical components of the supergeometric currents, i.e.,
degree zero components, in 
\eeqref{2dstandalekseevstrtoblsupercommutation1}--\eqref{2dstandalekseevstrtoblsupercommutation3}:
\beqa
    \{J_{0(f)}(\sigma),J_{0(g)}(\sigma^\prime)\}_{PB}&=&0, 
\\
    \{J_{1(X+\alpha)}(\sigma),J_{0(g)}(\sigma^\prime)\}_{PB}
&=& - \rho(X+\alpha) J_{0(g)}(x(\sigma)) 
\delta(\sigma-\sigma^\prime),
\\
\{J_{1(X+\alpha)}(\sigma),J_{1(Y+\beta)}(\sigma^\prime)\}_{PB}
  &=&
- J_{1 (\couranth{X+\alpha}{Y+\beta})}(\sigma)
\delta(\sigma-\sigma^\prime)
\nom\\
  &&+
\langle (X + \alpha), (Y + \beta) \rangle
(\sigma^\prime)
\partial_{\sigma} \delta(\sigma-\sigma^\prime).
\label{2dstandcurrentalgebra}
\eeqa
This coincides with the generalized current algebra described
in \cite{Alekseev:2004np}. 
\end{example}

\subsection{Poisson bracket twisted by $R$-flux}\label{PBtwitedR}
\noindent
By the same method as in the previous section, 
we derive a current algebra with $R$-flux
from the supergeometric data of the Poisson Courant algebroid.

Let us consider a two-dimensional worldsheet $\Sigma = S^1 \times \bR$
and take the tangent bundle $TLM = \Map(S^1, TM)$
of the loop space $LM= \Map(S^1, M)$ as phase space,
dual to the cotangent bundle $T^*LM = \Map(S^1, T^*M)$ 
in the standard Courant algebroid case.
Let $x^i(\sigma)$ be local coordinates on the base and 
$q^i(\sigma)$ local coordinates on the fiber,
where $\sigma$ parametrizes $S^1$.

We would like to construct Poisson brackets between 
$x^i(\sigma)$ and $q^i(\sigma)$. 
For that, we use the QP-manifold 
of the Poisson Courant algebroid,
which is $\calM = 
T^*[2]T^*[1]M$ with Q-structure
$\Theta = 
\pi^{ij}(x) \xi_i p_j
- \frac{1}{2} \frac{\partial \pi^{jk}}{\partial x^i}(x) q^i p_j p_k
+ \frac{1}{3!} R^{ijk}(x) p_i p_j p_k$.
We take a Lagrangian submanifold $\calL_0 = T[1]M$ of $\calM$
parametrized by local coordinates $(x^i, q^i)$.
The derived bracket on $T[1]M$ is
\begin{eqnarray}
\sbv{\sbv{x^i}{\Theta}}{x^j} &=& 0,
\label{contratargetcommu1}
\\
\sbv{\sbv{x^i}{\Theta}}{q^j} 
&=& 
\pi^{ij}(x),
\label{contratargetcommu2}
\\ 
\sbv{\sbv{q^i}{\Theta}}{q^j}
&=& 
- R^{ijk}(x) p_k
+ \frac{\partial \pi^{ij}}{\partial x^k}(x) q^k.
\label{contratargetcommu3}
\end{eqnarray}

The transgression 
of \eeqref{contratargetcommu1}--\eqref{contratargetcommu3} 
gives the derived brackets  on the superfields.
If we restrict them without twisting 
to the canonical Lagrangian submanifold parametrized by
$\bxi_i = \bp_i =0$,
we obtain the Poisson bracket without $R$-flux,
\begin{eqnarray}
\{\bx^i(\sigma, \theta),\bx^j(\sigma^{\prime}, \theta^{\prime})\}_{PB} 
&=& 0,
\label{contracommutationrelationssimple1}
\\
\{\bx^i(\sigma, \theta) ,\bq^j(\sigma^{\prime}, \theta^{\prime})\}_{PB} 
&=& 
- \pi^{ij}(\bx(\sigma, \theta))
\delta(\sigma-\sigma^\prime)
\delta(\theta-\theta^\prime), 
\label{contracommutationrelationssimple2}
\\ 
\{\bq^i(\sigma, \theta), \bq^j(\sigma^{\prime}, \theta^{\prime})\}_{PB}
&=& 
- \frac{\partial \pi^{ij}}{\partial \bx^k}(\bx(\sigma, \theta)) \bq^k
(\sigma, \theta)
\delta(\sigma-\sigma^\prime)
\delta(\theta-\theta^\prime).
\label{contracommutationrelationssimple3}
\end{eqnarray}
In order to introduce $R$-flux, 
we would like to consider a nontrivial restriction with twisting.

For simplicity, we assume that $\pi$ is nondegenerate.
In order to obtain an AS type current algebra,
we take
the Liouville $1$-form induced by 
the symplectic form $\bomega_{\calL_0}$
defined by the Poisson bracket
\eqref{contracommutationrelationssimple1}--\eqref{contracommutationrelationssimple3}.
This is $\alpha_0 = - \int_{\calX} \mu \ \bq^i (\pi^{-1})_{ij} \bbd \bx^j 
+ \cdots$, where $\cdots$ contains terms without $\bq^i$.

Twisting by $\alpha_0$ induces the twist
$\bp_i \rightarrow \bp_i - (\pi^{-1})_{ij} \bbd \bx^j$.
After the restriction $\bxi_i=\bp_i = 0$ to the canonical Lagrangian 
submanifold,
we get the Poisson brackets with $R$-flux,
\begin{eqnarray}
\{\bx^i(\sigma, \theta),\bx^j(\sigma^{\prime}, \theta^{\prime})\}_{PB} 
&=& 
0,
\label{supercommurelationR1}
\\
\{\bx^i(\sigma, \theta) ,\bq^j(\sigma^{\prime}, \theta^{\prime})\}_{PB} 
&=& 
- \pi^{ij}(\bx(\sigma, \theta))
\delta(\sigma-\sigma^\prime)
\delta(\theta-\theta^\prime), 
\label{supercommurelationR2}
\\ 
\{\bq^i(\sigma, \theta), \bq^j(\sigma^{\prime}, \theta^{\prime})\}_{PB}
&=& 
- \left(R^{ijk}(\bx) (\pi^{-1})_{kl} \bbd \bx^l
+ \frac{\partial \pi^{ij}}{\partial \bx^k}(\bx) \bq^k
\right)
\delta(\sigma-\sigma^\prime)
\delta(\theta-\theta^\prime).~~~~~~~~~\label{supercommurelationR3}
\end{eqnarray}

Physical Poisson brackets are the degree zero components of 
these equations.
Here we denote physical fields by $x^i(\sigma) = x^{(0)i}(\sigma)$
and $q^i(\sigma) = q^{(1)i}(\sigma)$.
Then, the Poisson brackets on the physical canonical quantities 
are
\begin{eqnarray}
\{x^i(\sigma),x^j(\sigma^{\prime})\}_{PB} 
&=& 0,
\label{2dcanoicalcommutation1}
\\
\{x^i(\sigma) ,q^j(\sigma^{\prime})\}_{PB} 
&=& 
\pi^{ij}(x(\sigma))
\delta(\sigma-\sigma^\prime),
\label{2dcanoicalcommutation2}
\\ 
\{q^i(\sigma), q^j(\sigma^{\prime})\}_{PB}
&=& 
\left(- R^{ijk}(x) (\pi^{-1})_{kl}(x) \partial_{\sigma} x^l
+ \frac{\partial \pi^{ij}}{\partial x^k}(x) q^k
\right)
(\sigma)
\delta(\sigma-\sigma^\prime).
\label{2dcanoicalcommutation3}
\end{eqnarray}

The relations
\eeqref{2dcanoicalcommutation1}--(\ref{2dcanoicalcommutation3})
can also be derived by $\beta$-transformation.
If $R=0$, the relations
\eqref{2dcanoicalcommutation1}--\eqref{2dcanoicalcommutation3}
are obtained by pullback 
of the Poisson structure on $TM$,
lifted from the Poisson structure $\pi$ on $M$ \cite{GU1},
to the mapping space.
In this case, the $R$-term is introduced by $\beta$-transformation, 
$x^i \rightarrow x^i$ and 
$q^i \rightarrow q^i + \beta^{ij} \pi^{-1}_{jk} \partial_{\sigma} x^k$,
where 
$\beta = \frac{1}{2} 
\beta^{ij}(x) \frac{\partial}{\partial x^i} \wedge 
\frac{\partial}{\partial x^j}$ is a bivector field such that 
$[\pi, \beta]_S =R$.

The symplectic form 
of Alekseev-Strobl type
which induces
(\ref{2dcanoicalcommutation1})--(\ref{2dcanoicalcommutation3})
is
\beq
    \omega=
\int_{S^1}d\sigma \ (\pi^{-1})_{ij} \delta x^i \wedge \delta q^j 
-
\frac{1}{2} \int_{S^1} d\sigma 
\left(- R^{ijk} 
(\pi^{-1})_{kl} \partial_{\sigma} x^l
+ \frac{\partial \pi^{ij}}{\partial x^k} q^k \right)
(\pi^{-1})_{im} \delta x^m \wedge 
(\pi^{-1})_{jn} \delta x^n.
\label{2dsymplectic}
\eeq

\subsection{Contravariant current algebras with $R$-flux}
\label{contracurrentwithR}
\noindent
Here, currents are constructed from functions 
of degree equal to or less than one
on the target space,
$C_0 \oplus C_1 = \{f \in C^{\infty}(T^*[2]T[1]M) | |f| \leq 1 \}$, 
which is the same space as in the case of 
the AS current algebra.
Take a function of degree zero $j_0 = f(x)$
and a function of degree one $j_1 = X^i(x) p_i + \alpha_i(x) q^i$.
By transgression of 
$j_0$ and $j_1$ to the mapping space,
twisting by $\alpha_0$
such that $\bp_i \rightarrow \bp_i - (\pi^{-1})_{ij} \bbd \bx^j$,
and restricting them to the canonical Lagrangian submanifold,
we obtain the supergeometric currents,
\begin{eqnarray}
&& \bJ_{(0)(f)}(\epsilon_{(1)})
= \tilpi_* e^{\delta_{\alpha_0}} \mu_* \epsilon_{(1)} \ev^*  j_{(0)(f)}
= \int_{T[1]S^1} \mu \epsilon_{(1)} f(\bx),
\nonumber \\
&&
\bJ_{(1)(X+ \alpha)}(\epsilon_{(0)})
= \tilpi_* e^{\delta_{\alpha_0}} \mu_* \epsilon_{(0)} \ev^* j_{(1)(u,\alpha)}
= \int_{T[1]S^1} \mu \epsilon_{(0)} 
(- X^i(\bx) (\pi^{-1})_{ij} \bbd \bx^j + \alpha_i(\bx) \bq^i).
\nonumber
\end{eqnarray}
If we take the degree zero components of the superfields, we obtain
AS type currents,
\beq
    J_{0(f)}(\sigma)=f(x(\sigma)), \quad
    J_{1(X+\alpha)}(\sigma)
     =
X^i(x(\sigma)) (\pi^{-1})_{ij} \partial_{\sigma} x^j(\sigma)
+ \alpha_i(x(\sigma)) q^i(\sigma).
\label{2dcurrent}
\eeq
The algebra of these supergeometric currents
is computed from the Poisson brackets of the canonical quantities
(\ref{supercommurelationR1})--(\ref{supercommurelationR3}):
\beqa
    \{\bJ_{0(f)}(\epsilon),
\bJ_{0(g)}(\epsilon^{\prime})
\}_{PB}
&=&
0,
\nom \\
     \{\bJ_{1(X+\alpha)}(\epsilon),
\bJ_{0(g)}(\epsilon^{\prime})\}_{PB}
&=&
\rho(X + \alpha) \bJ_{0(g)}
(\epsilon \epsilon^{\prime}),\nom\\
    \{\bJ_{1(X+\alpha)}(\epsilon),
\bJ_{1(Y+\beta)}(\epsilon^{\prime})\}_{PB}
    &=&
\bJ_{1 (\courantr{X+\alpha}{Y+\beta}}(\epsilon \epsilon^{\prime})
\nonumber \\
&&
+ \int_{T[1]S^1} \mu \bbd \epsilon_{(0)} \epsilon^{\prime}_{(0)}
\bracket{X + \alpha}{Y + \beta}(\bx),
\label{2dcontraalekseevstrtoblsupercommutation}
\eeqa
where
\beqa
    \courantr{X+\alpha}{Y+\beta}
&=& [\alpha, \beta]_{\pi} +
L_{\alpha}^{\pi}Y - \iota_{\beta} d_{\pi} X
- R(\alpha, \beta, -),
\eeqa
is the contravariant Dorfman bracket with $R$-flux on $TM \oplus T^*M$
and 
$\rho(X+\alpha) = \pi^{\sharp} (\alpha)$
is the anchor map.
Component expansions give rise to 
physical current algebras:
\begin{eqnarray}
    \{J_{0(f)}(\sigma),J_{0(g)}(\sigma^\prime)\}_{PB}&=&0, 
\label{2dRcurrentalgebra1}
\\
    \{J_{1(X+\alpha)}(\sigma),J_{0(g)}(\sigma^\prime)\}_{PB}
&=& - \rho(X+\alpha) J_{0(g)}(x(\sigma)) 
\delta(\sigma-\sigma^\prime),
\label{2dRcurrentalgebra2} \\
\{J_{1(X+\alpha)}(\sigma),J_{1(Y+\beta)}(\sigma^\prime)\}_{PB}
  &=&
- J_{1(\courantr{X+\alpha}{Y+\beta})}(\sigma)
\delta(\sigma-\sigma^\prime)
\nom\\
  &&+
\langle X + \alpha, Y+ \beta \rangle
(\sigma^\prime)
\partial_{\sigma} \delta(\sigma-\sigma^\prime).
\label{2dRcurrentalgebra3}
\end{eqnarray}
This formula (\ref{2dRcurrentalgebra1})--(\ref{2dRcurrentalgebra3}) 
is consistent, even if
the Poisson structure $\pi$ is degenerate.
Therefore, we do not need to impose a nondegeneracy condition
for $\pi$ in the current algebra.

\section{Conclusions and discussion}
\noindent
The Poisson Courant algebroid, which is a contravariant object 
of the standard Courant algebroid, 
has been formulated by supergeometric construction.
The duality between these two specific Courant algebroids 
has been analyzed in detail.
As a result,
the duality transformation is
a canonical transformation on the graded symplectic manifold
and the 
transformation
between the $3$-form $H$-flux in the standard Courant algebroid 
and the trivector $R$-flux in the Poisson Courant algebroid 
has been derived.
In \cite{Hofman:2002rv, Klimcik:2001vg, Severa:2001qm},
twisting of a bivector field by a $3$-form $H$, 
a so-called twisted Poisson structure, 
has been discussed.
From the above duality, we have obtained its contravariant geometric 
structure in \eeqref{twisted2formbyR}, 
twisting of a $2$-form by a trivector field $R$.

Moreover, we have shown that this duality is, from the mathematical viewpoint,
the generalization of the correspondence between 
the de Rham cohomology and the Poisson cohomology. 
We also discussed that the same duality can be derived on the sigma model level.

By using the supergeometric formulation, we have constructed a 3-dimensional AKSZ sigma model and a 2-dimensional boundary sigma model with the structure of a Poisson Courant algebroid. 
From the physical viewpoint, we are considering a theory of a topological membrane on a Poisson manifold. From the general form of the homological function given in (\ref{GeneralHomologicalFunction}), (see also \cite{Hofman:2002rv,Ikeda:2002wh}) 
we can introduce all types of third rank tensors ($H$,$F$,$Q$ and $R$) with various covariant and contravariant suffixes. However, it is remarkable that the 3-vector $R$ can only be introduced on the Poisson manifold.
Then, we derived the topological string on the Poisson manifold as the boundary theory of that topological membrane. Of course, the theory obtained in this way can be identified with the Poisson sigma model by field redefinition. 
What we found is that there is a specific way to lift the boundary theory to the topological membrane including $R$-flux on the Poisson manifold.

We have also constructed a current algebra with $R$-flux on the tangent space of the loop space from the target space QP-manifold data.
The resulting current algebra is the contravariant counterpart 
of the current algebra with $H$-flux of Alekseev-Strobl type.


{The $R$-flux has also been discussed in 
\cite{
Aldazabal:2011nj,Andriot:2012an}
 using double field theory. 
There, the nongeometric $R$-flux is characterized as a Jacobiator, the quantity corresponding to the anomaly of the Jacobi identity, i.e.
$R^{ijk} \sim \beta^{l[i} \frac{\partial \beta^{jk]}}{\partial x^l}$.
In section \ref{PCAfromDG}, we have discussed the Poisson Courant algebroid
and its trivector field $R$
from the point of view of double field theory.
If we take the special solution of the section condition 
defined by the Poisson structure $\pi$, the resulting spacetime
has the Poisson Courant algebroid structure.
Therefore, we have found that our formalism describes 
the $R$-flux in the frame specified by this particular solution.
}

In our formulation, $\beta$ is independent of the Poisson bivector $\pi$,
thus we can consider the special case
$[\pi + \beta, \pi + \beta]_S=0$.
It means that $\pi + \beta$ is again a Poisson structure.
Note that it does not mean a deformation
of the Poisson Courant algebroid.
This is a Maurer-Cartan condition of $\beta$,
$d_{\pi} \beta + \frac{1}{2} [\beta, \beta]_S =0$,
and we obtain the $R$-flux as the Jacobiator
\cite{Blumenhagen:2010hj, Blumenhagen:2011ph},
\begin{eqnarray}
&& R = - \frac{1}{2} [\beta, \beta]_S.
\end{eqnarray}
This is the same formula in the definition inspired by the 
double field theory.
The meaning of this observation will be discussed in future work.


\renewcommand{\theequation}{A.\arabic{equation}}
\setcounter{equation}{0}
\appendix

\section{Formulas in graded differential calculus}
\noindent
We summarize formulas of graded symplectic geometry.

\subsection{Basic definitions}
\noindent
Let $z$ be a local coordinate on a graded manifold $\calM$.
A differential on a function is defined by
\begin{eqnarray}
d f(z) &=& d z^a \frac{\ld f}{\partial z^a}.
\end{eqnarray}
A vector field $X$ is expanded using local coordinates by
\begin{eqnarray}
X = X^a(z) \frac{\ld}{\partial z^a}.
\end{eqnarray}
The interior product is defined by the differentiation by
the following graded vector field on $T[1]\calM$,
\begin{eqnarray}
\iota_X &=& (-1)^{|X|}
X^a(z) \frac{\ld}{\partial d z^a},
\end{eqnarray}
where we define
$\frac{\ld}{\partial d z^a} d z^b = \delta^b{}_a$.
For a graded differential form $\alpha$, we denote $|\alpha|$ as total degree 
(form degree plus degree by grading) of $\alpha$.
Note that $|d| = 1$, $|d z^a| = |z^a| +1$
and
$|\iota_X|= |X| -1$.
%
%
For vector fields, $X = X^a(z) \frac{\ld}{\partial z^a}, 
Y = Y^a(z) \frac{\ld}{\partial z^a}$,
the graded Lie bracket is
\begin{eqnarray}
[X, Y] 
&=& X^a \frac{\ld Y^b}{\partial z^a} \frac{\ld}{\partial z^b}
- (-1)^{|X||Y|} Y^a \frac{\ld X^b}{\partial z^a} \frac{\ld}{\partial z^b}.
\end{eqnarray}
We obtain the following formula,
\begin{eqnarray}
Xf &=& (-1)^{|X|} \iota_X df
= (-1)^{(|f|+1)|X|} df(X),
\label{differentialoffunction}
\end{eqnarray}
where
\begin{eqnarray}
d z^a \left(\frac{\ld}{\partial z^b} \right) &=& \delta^a{}_b.
\end{eqnarray}
\begin{proof}
We prove \eeqref{differentialoffunction}.
Since $Xf = X^a(z) \frac{\ld f}{\partial z^a}$,
we have
\begin{eqnarray}
(-1)^{|X|} \iota_X df
&=& (-1)^{|X|} (-1)^{|X|} X^a(z) 
\frac{\ld}{\partial d z^a} \left(
d z^a \frac{\ld f}{\partial z^a} \right).
\end{eqnarray}
Therefore,
\begin{eqnarray}
df(X)
&=&
d z^a \frac{\ld f}{\partial z^a}
\left(
X^b(z) \frac{\ld}{\partial z^b}
\right)
\nonumber \\
&=&
(-1)^{(|f|-|z|)|X|} 
\left[
d z^a \left(X^b(z)
\frac{\ld}{\partial z^b}
\right) 
\right]
\frac{\ld f}{\partial z^a}
\nonumber \\
&=&
(-1)^{(|f|-|z|)|X|} 
(-1)^{(|X| - |z|)(|z|+1)} 
X^b(z) 
\left[
d z^a \left(
\frac{\ld}{\partial z^b}
\right)
\right]
\frac{\ld f}{\partial z^a}
\nonumber \\
&=&
(-1)^{(|f|+1)|X|} 
X^a(z) \frac{\ld f}{\partial z^a}.
\end{eqnarray}
\hfill\qed
\end{proof}

\subsection{Cartan formulas}
\noindent
The Lie derivative is defined by
\begin{eqnarray}
L_X = \iota_X d - (-1)^{(|X|-1) \times 1} d \iota_X
= \iota_X d + (-1)^{|X|} d \iota_X.
\end{eqnarray}
Its degree is $|L_X| = |X|$.

Let $\alpha$ and $\beta$ be graded differential forms.
We can show the following graded Cartan formulas,
\begin{eqnarray}
\alpha \wedge \beta &=& (-)^{|\alpha||\beta|}\beta \wedge \alpha,
\\
d (\alpha \wedge \beta) &=& 
d \alpha \wedge \beta + (-1)^{|\alpha|} \alpha \wedge d \beta,
\\
\iota_X (\alpha \wedge \beta) &=& 
\iota_X \alpha \wedge \beta + (-1)^{|\alpha|(|X|+1)} 
\alpha \wedge \iota_X \beta,
\\
L_X (\alpha \wedge \beta) &=& 
L_X  \alpha \wedge \beta + (-1)^{|\alpha||X|} 
\alpha \wedge L_X \beta,
\\
L_X d &=& (-1)^{|X|}  d L_X,
\\
\iota_X \iota_Y - (-1)^{(|X|- 1)(|Y|-1)} \iota_Y \iota_X &=& 0,
\\
L_X \iota_Y - (-1)^{|X|(|Y|- 1)} \iota_Y L_X &=& \iota_{[X, Y]},
\\
L_X L_Y - (-1)^{|X||Y|} L_Y L_X &=& L_{[X, Y]}.
\end{eqnarray}

\subsection{Differential forms}
\noindent
Let $\alpha = d z^{a_1} \wedge \cdots \wedge d z^{a_m} \alpha_{a_1 \cdots a_m}(z)$
be an $m$-form on $\mathcal{M}$.
The contraction of $\alpha(X, -, \cdots, -)$ with a vector field $X$ on $\mathcal{M}$ is
\begin{eqnarray}
\alpha(X, -, \cdots, -)
= (-1)^{|X|(|\alpha|+1)} \iota_X \alpha(-, \cdots, -).
\label{interiorderivativeondform}
\end{eqnarray}

\begin{proof}
\begin{eqnarray}
\alpha(X, -, \cdots, -)
&=&
d z^{a_1} \wedge \cdots \wedge
d z^{a_m} \alpha_{a_1 \cdots a_m}(z)
\left (X^b 
\frac{\ld}{\partial z^b}\right)
\nonumber \\
&=&
(-1)^{|X|(|\alpha|-|z|-1)} 
d z^{a_1} 
\left(X^b
\frac{\ld}{\partial z^b}
\right) 
d z^{a_2} \wedge \cdots \wedge
d z^{a_m} \alpha_{a_1 \cdots a_m}(z)
\nonumber \\
&=&
(-1)^{|X|(|\alpha|-|z|-1)} 
(-1)^{(|X| - |z|)(|z|+1)} 
X^{a_1} 
d z^{a_2} \wedge \cdots \wedge
d z^{a_m} \alpha_{a_1 \cdots a_m}(z)
\nonumber \\
&=&
(-1)^{|X||\alpha|} 
X^{a_1} 
d z^{a_2} \wedge \cdots \wedge
d z^{a_m} \alpha_{a_1 \cdots a_m}(z)
\nonumber \\
&=&
(-1)^{|X||\alpha|} 
(-1)^{|X|}
\iota_X \alpha.
\end{eqnarray}
\hfill\qed
\end{proof}

By induction 
using \eeqref{interiorderivativeondform}, 
we obtain the general formula,
\begin{eqnarray}
\alpha(X_m, X_{m-1}, \cdots, X_1)
&=& - (-1)^{\sum_{i=1}^m |X_i|(|\alpha|+i)} \iota_{X_m} 
\cdots \iota_{X_1} \alpha,
\\
%
%
\alpha(X_m, \cdots, X_j, \cdots, X_i, \cdots X_1)
&=& - (-1)^{|X_i||X_j|} 
\alpha(X_m, \cdots, X_i, \cdots, X_j, \cdots X_1).
\end{eqnarray}
Especially, if $\alpha$ is a $2$-form, we derive
\begin{eqnarray}
\alpha(X, Y)
&=& - (-1)^{|X||Y|} \alpha(Y, X).
\end{eqnarray}

\subsubsection{Exterior derivatives}
\noindent
The exterior derivative on a function is given by \eeqref{differentialoffunction},
\begin{eqnarray}
df(X)
= (-1)^{|X|(|f|+1)} Xf.
\end{eqnarray}
Let $\alpha$ be a $1$-form on $\mathcal{M}$.
Then, from the Cartan formulas, we obtain
\begin{eqnarray}
d \alpha(X_1, X_2)
&=&
(-1)^{|X_1||\alpha|} X_1 \alpha(X_2)
- (-1)^{|X_2||\alpha|} (-1)^{|X_1||X_2|} X_2 \alpha(X_1)
- \alpha([X_1, X_2]).
\end{eqnarray}
For a $2$-form $\alpha$, the formula gives
\begin{eqnarray}
d \alpha(X_1, X_2, X_3)
&=&
(-1)^{|X_1|(|\alpha| +1)} X_1 \alpha(X_2, X_3)
- (-1)^{|X_2|(|\alpha| +1)} (-1)^{|X_1||X_2|} X_2 \alpha(X_1, X_3)
\nonumber \\ &&
+ (-1)^{|X_3|(|\alpha| +1)} (-1)^{(|X_1|+|X_2|)|X_3|} X_3 \alpha(X_1, X_2)
- \alpha([X_1, X_2], X_3)
\nonumber \\ &&
+ (-1)^{|X_2||X_3|} \alpha([X_1, X_3], X_2)
- (-1)^{|X_1|(|X_2| + |X_3|)} \alpha([X_2, X_3], X_1).
\end{eqnarray}
Let $\alpha = d z^{a_1} \wedge \cdots \wedge d z^{a_m} \alpha_{a_1 \cdots a_m}(z)$
be an $m$-form on $\mathcal{M}$. 
Then, we can prove the following formula by induction,
\begin{eqnarray}
&& d \alpha(X_1, X_2, \cdots, X_m)
\nonumber \\
&& = 
\sum_{i=1}^m (-1)^{i-1} (-1)^{|X_i|(|\alpha|+m)} 
(-1)^{\sum_{k=1}^{i-1} |X_i||X_k|}
X_i \alpha(X_1, \cdots, \hat{X_i}, \cdots, X_m)
\nonumber \\
&& 
+
\sum_{i < j}
(-1)^{i+j} 
(-1)^{\sum_{k=1}^{i-1} |X_i||X_k|
+ \sum_{l=1, l \neq j}^{j-1} |X_j||X_l|}
\alpha([X_i, X_j], 
\cdots, \hat{X_i}, \cdots, \hat{X_j}, \cdots, X_m).
\end{eqnarray}

\subsection{Graded symplectic form and Poisson bracket}
\noindent
Let $\omega$ be a symplectic form of degree $n$.
Since $\omega$ is a $2$-form, its total degree is $|\omega| = n+2$.
Let $z = (q^a, p_a)$ be 
Darboux coordinates such that 
$|q| + |p| =n$. Then we obtain
\begin{eqnarray}
\omega &=& (-1)^{|q|(|p|+1)} d q^a \wedge d p_a 
= (-1)^{n|q|} d q^a \wedge d p_a 
\nonumber \\
&=& (-1)^{n|q|}(-1)^{(|q|+1)(|p|+1)} d p_a \wedge d q^a
= (-1)^{|p|+1} d p_a \wedge d q^a.
\end{eqnarray}
Then, the Liouville $1$-form $\omega = - d \vartheta$ is given by
\begin{eqnarray}
\vartheta 
&=& 
(-1)^{|p|} p_a d q^a
= - (-1)^{n +1 - |q|} p_a d q^a
= (-1)^{|q||p|} d q^a p_a
\\
&=& - (-1)^{|q|(|p|+1)} q^a d p_a
= - d p_a q^a.
\end{eqnarray}

The Hamiltonian vector field $X_f$ of a function $f$ is defined by
\begin{eqnarray}
\iota_{X_f} \omega &=& - d f.
\end{eqnarray}
Its total degree is $|X_f| = |f|-n$.
In order to obtain the Darboux coordinate expression of $X_f$,
assume that $X = X_a \frac{\ld }{\partial p_a}
+
Y^a \frac{\ld }{\partial q^a}$.
Then we derive
\begin{eqnarray}
\iota_{X_f} \omega &=& 
 \left(
(-1)^{|X| + p} X_a \frac{\ld }{\partial d p_a}
+ 
(-1)^{|X| + q} 
Y^a \frac{\ld }{\partial d q^a}
\right)
\cdot \left( (-1)^{n|q|} d q^a \wedge d p_a \right)
\nonumber \\
&=& - dq^a \frac{\ld f}{\partial q^a}
- dp_a \frac{\ld f}{\partial p_a}.
\end{eqnarray}
By solving this equation, we finally obtain
\begin{eqnarray}
X_f &=& 
\frac{f \rd}{\partial q^a}\frac{\ld }{\partial p_a}
- (-1)^{|q||p|}
\frac{f \rd}{\partial p_a}\frac{\ld }{\partial q^a}.
\end{eqnarray}
Here, $\frac{f \rd}{\partial q^a} = 
(-1)^{(|f|-q)q} \frac{\ld f}{\partial q^a}$
is the right derivative.

The graded Poisson bracket is defined by
\begin{eqnarray}
\sbv{f}{g} &=& X_f g
= (-1)^{|f|+n} \iota_{X_f} dg
= (-1)^{|f|+n + 1} \iota_{X_f} \iota_{X_g} \omega.
\end{eqnarray}
It satisfies
\begin{eqnarray*}
\sbv{f}{g}&=&-(-1)^{(|f| - n)(|g| - n)} \sbv{g}{f},\\
\sbv{f}{g  h}&=&\sbv{f}{g} h
+ (-1)^{(|f| - n)|g|} g \sbv{f}{h},\\
\{f,\{g,h\}\}&=&\{\{f,g\},h\}+(-1)^{(|f|-n)(|g|-n)}\{g,\{f,h\}\}.
\end{eqnarray*}
For the Darboux coordinates, we get the relations 
\begin{eqnarray}
\sbv{q^a}{p_b} &=& \delta^a{}_b,
\qquad
\sbv{p_b}{q^a} = - (-1)^{|q||p|}\delta^a{}_b.
\end{eqnarray}
For functions, $f = f(q, p)$ and $g = g(q, p)$,
the graded Poisson bracket is given by
\begin{eqnarray}
\sbv{f}{g} &=& 
\frac{f \rd}{\partial q^a}\frac{\ld g}{\partial p_a}
- (-1)^{|q||p|}
\frac{f \rd}{\partial p_a}\frac{\ld g}{\partial q^a}.
\end{eqnarray}


$X$ is called \textit{symplectic vector field}, if $L_X \omega = 0$, 
i.e., $d \iota_X \omega =0$.
Let $X, Y$ be symplectic vector fields. 
Then, $[X,Y]$ is the Hamiltonian vector field for 
$ - (-1)^{|X|}\iota_X \iota_Y \omega$.
\begin{proof}
\begin{eqnarray}
\iota_{[X, Y]} \omega &=& 
(L_X \iota_Y - (-1)^{|X|(|Y|-1)} \iota_Y L_X ) \omega
= (-1)^{|X|} d \iota_X \iota_Y \omega
\nonumber \\
&=& - d [- (-1)^{|X|} \iota_X \iota_Y \omega].
\end{eqnarray}
\hfill\qed
\end{proof}
If $X = X_f$, $Y = X_g$ are Hamiltonian vector fields, then the following equation holds, 
\begin{eqnarray}
\iota_{[X_f, X_g]} \omega &=& 
(-1)^{|f| + n} d \iota_{X_f} \iota_{X_g} \omega.
\end{eqnarray}
Therefore, we get
\begin{eqnarray}
X_{\sbv{f}{g}}&=&  - [X_f, X_g].
\end{eqnarray}
Since $\iota_{X_f} \iota_{X_g} \omega
= - (-1)^{|f|n + |g|(n+1)} \omega(X_g, X_f)$, we easily derive
\begin{eqnarray}
\sbv{f}{g} &=& (-1)^{|f| +n + 1} \iota_{X_f} \iota_{X_g} \omega
\nonumber \\
&=&
(-1)^{(|f|+ |g|)(n+1)} \omega(X_g, X_f)
=
(-1)^{|f||g| + n+1} \omega(X_f, X_g).
\end{eqnarray}

We consider the AKSZ construction on $\Map(\calX, \calM)$.
Let $D$ be a differential on $\calX$. It can be locally expressed as
$D = \theta^{\mu}\frac{\partial}{\partial \sigma^{\mu}}$.
We denote with $\hat{D}$ the vector field on $\Map(\calX, \calM)$
of degree $1$, which is induced by $D$.
Then the following equation holds,
\begin{eqnarray}
\sbv{\iota_{\hat{D}} \mu_* \ev^* \vartheta}{\mu_* \ev^* f} 
&=& - \iota_{\hat{D}} \mu_* \ev^* d f
\left(= \int d^{n+1}\sigma d^{n+1}\theta \bbd f(\sigma, \theta) \right),
\end{eqnarray}
for $f \in C^{\infty}(\calM)$.
\begin{proof}
$S_0 = \iota_{\hat{D}} \mu_* \ev^* \vartheta$ is a Hamiltonian 
for the vector field $\hat{D}$,i.e.,
$X_{S_0} = \hat{D}$. Therefore, we get
\begin{eqnarray}
\sbv{\iota_{\hat{D}} \mu_* \ev^* \vartheta}{\mu_* \ev^* f} 
&=& 
\sbv{S_0}{\mu_* \ev^* f} 
\nonumber \\ &=&
(-1)^{|S_0|} \iota_{\hat{D}} \iota_{X_{\mu_* \ev^* f}} \bomega
\nonumber \\ &=&
- \iota_{\hat{D}} \mu_* \ev^* d f.
\end{eqnarray}

\end{proof}

\section{Formulas on the mapping space}
\noindent

\subsection{Functional differential calculus}
\noindent
We list the formulas, that we use on the mapping space $\Map(\calX, \calM)$. 
Let $X$ be a manifold of dimension $d = n+1$. The mapping space functions are superfields. They depend on variables on $\calX = T[1]X$, 
which is a $(d,d)$-dimensional supermanifold with even local coordinates 
$\sigma^{\mu}$ and odd local coordinates $\theta^{\mu}$, 
where $\mu=1,\dots, d$.

A component expansion of 
a superfield $\Phi(\sigma,\theta)$ of degree $|\Phi|$ 
in Grassmann variables is defined as
\begin{equation}
\Phi(\sigma,\theta) 
= \sum_{j=0}^{d}\frac{1}{j!}\phi_{\mu_{1}\cdots\mu_{j}}(\sigma) \theta^{\mu_{1}}\cdots\theta^{\mu_{j}},
\end{equation}
where the $j=0$ term $\phi(\sigma)$
is not accompanied by $\theta^{\mu}$.
Since $|\theta^{\mu}| = 1$, we get $|\phi_{\mu_{1}\cdots\mu_{j}}| = |\Phi| - j$.

The functional derivative on the mapping space is 
\begin{equation}
	\frac{\overrightarrow{\delta} \Phi(\sigma,\theta)}{\delta \Phi(\sigma',\theta')} = \delta^{d}(\sigma' - \sigma)\delta^{d}(\theta' - \theta).
\end{equation}
Expanding this equation in components, 
we obtain the formula for the left functional derivative,
\begin{equation}
\frac{\overrightarrow{\delta}}{\delta \Phi(\sigma,\theta)} 
= \sum_{j=0}^{d} \frac{(-1)^{d-j}}{j!(d-j)!}
\theta^{\mu_{1}}\cdots\theta^{\mu_{j}}\epsilon_{\mu_{1}\cdots\mu_{j}\mu_{j+1}\cdots\mu_{d}}
\frac{\overrightarrow{\delta}}{\delta\phi_{\mu_{j+1}\cdots\mu_{d}}(\sigma)},
\end{equation}
where $\epsilon_{\mu_{1}\cdots\mu_{j}\mu_{j+1}\cdots\mu_{d}}$ is the 
completely antisymmetric Levi-Civita symbol.
By degree counting, we obtain 
$\left|\frac{\overrightarrow{\delta}}
{\delta\phi_{\mu_{j+1}\cdots\mu_{d}}(\sigma)}\right| = -(|\Phi| - d + j)$.
We require the following identity for the right functional derivative,
\begin{equation}
\frac{\overrightarrow{\delta} F}{\delta\Phi} 
= (-1)^{|F|(|\Phi| - d)}\frac{F\overleftarrow{\delta}}{\delta\Phi},
\end{equation}
for an arbitrary superfield $F$.
From this equation, we have the right derivative
for components,
\begin{equation}
\frac{\overleftarrow{\delta}}{\delta\Phi(\sigma,\theta)} 
= \sum_{j=0}^{d}\frac{1}{j!(d-j)!}(-1)^{|\Phi| + j(|\Phi| + d + 1)}
\frac{\overleftarrow{\delta}}{\delta\phi_{\mu_{j+1}\cdots\mu_{d}}(\sigma)}
\theta^{\mu_{1}}\cdots\theta^{\mu_{j}}
\epsilon_{\mu_{1}\cdots\mu_{j}\mu_{j+1}\cdots\mu_{d}}.
\end{equation}
Computing the right derivative
$\frac{\Phi(\sigma,\theta)\overleftarrow{\delta}}{\delta\Phi(\sigma',\theta')}$
using the above formula,
we summarize
\begin{align}
\frac{\overrightarrow{\delta}\Phi(\sigma,\theta)}{\delta\Phi(\sigma',\theta')} &= \delta^{d}(\sigma' - \sigma)\delta^{d}(\theta' - \theta), \\
\frac{\Phi(\sigma,\theta)\overleftarrow{\delta}}{\delta\Phi(\sigma',\theta')} &= (-1)^{|\Phi|(1+d)+d}\delta^{d}(\sigma - \sigma')\delta^{d}(\theta - \theta').
\end{align}
The degrees of the right and left derivatives are
\begin{equation}
	\left|\frac{\overrightarrow{\delta}}{\delta\Phi(\sigma,\theta)}\right| = \left|\frac{\overleftarrow{\delta}}{\delta\Phi(\sigma,\theta)}\right| = d - |\Phi|.
\end{equation}
For an arbitrary superfield $F$, we have
\begin{equation}
	\left|\frac{\overrightarrow{\delta}F}{\delta\Phi(\sigma,\theta)}\right| = \left|\frac{F\overleftarrow{\delta}}{\delta\Phi(\sigma,\theta)}\right| = |F| + d - |\Phi|.
\end{equation}
%
%
The following left and right Leibniz rules hold for arbitrary superfields 
$F$ and $G$:
\begin{align}
	\frac{\overrightarrow{\delta}}{\delta\Phi}(FG) &= \frac{\overrightarrow{\delta}F}{\delta\Phi}\cdot G + (-1)^{|F|(d - |\Phi|)}F\cdot \frac{\overrightarrow{\delta} G}{\delta\Phi}, \\
	(FG)\frac{\overleftarrow{\delta}}{\delta\Phi} &=  F\cdot \frac{G\overleftarrow{\delta}}{\delta\Phi} + (-1)^{|G|(d - |\Phi|)}\frac{F\overleftarrow{\delta}}{\delta\Phi}\cdot G.
\end{align}
The measure on the worldvolume supermanifold is defined by
$\mu=d\sigma^{1}\cdots d\sigma^{d}d\theta^{d}\cdots d\theta^{1}$
and its degree is $|\mu| = -d$.
%
The following equation for the Grassmann delta function holds
\begin{equation}
	\int\mu_{\theta} \delta^{d}(\theta - \theta') \Phi(\sigma,\theta) = \Phi(\sigma,\theta')
\end{equation}
where $\mu_{\theta} = d^{d}\theta\cdots d^{1}\theta$.

\subsection{Graded symplectic geometry}
\noindent
In this subsection, we map
the structures on $\calM$ to structures on 
the target space $\Map(\calX, \calM)$ by the transgression map $\mu_* \ev^*$.

Let $\bbz^{i}(\sigma,\theta)$ be a local basis superfield of the mapping space 
$\Map(\calX, \calM)$, corresponding to a local coordinate $z^i$ on $\calM$.
We write a \textsl{vector field} on the mapping space for 
$X = X^i(z) \frac{\ld}{\partial z^i}$
as
\begin{equation}
X = \int_{\calX} \mu \ (-1)^{d|X^{i}|}X^{i}(\bbz(\sigma,\theta))
\frac{\overrightarrow{\delta}}{\delta \bbz^{i}(\sigma,\theta)}.
\end{equation}
Then, the \textsl{interior product} is 
\begin{equation}
\iota_{X} = (-1)^{|X|} \int_{\calX}\mu \ (-1)^{d|X^{i}|} 
X^{i}(\bbz(\sigma,\theta))
\frac{\overrightarrow{\delta}}{\delta(\delta \bbz^{i})(\sigma,\theta)}.
\end{equation}
The symplectic form $\bomega$ on the mapping space corresponding 
to $\omega = (-1)^{n|q|}\delta q^{i} \wedge \delta p_{i}$
is defined as
\begin{equation}
\bomega = \int_{\calX} \mu \ (-1)^{(d-1)|q|}\delta \bbq^{i}(\sigma,\theta)\wedge \delta \bbp_{i}(\sigma,\theta).
\end{equation}
We have $|\bomega| = |\mu| + 1 + |\bbq^{i}| + 1 + |\bbp_{i}| 
= -d + 2 + d - 1 = 1$ \footnote{This is degree counting to determine 
the sign factor. In fact, $\bomega$ is a $2$-form of degree $d-1$. Then
we count the sign factor as degree $d+1$.}.

The \textsl{differential on a function} $f$ is 
\begin{equation}
	\mbox{\boldmath $\delta$} f = \int_{\calX}\mu \ (-1)^{d(|\bbz^{i}|+1)}(\delta \bbz^{i})(\sigma,\theta) \frac{\overrightarrow{\delta}f}{\delta \bbz^{i}(\sigma,\theta)}.
\end{equation}
We define the \textsl{Liouville $1$-form} (the canonical $1$-form) 
$\mbox{\boldmath $\vartheta$}$ on the mapping space as
\begin{equation}
	\bomega = -\mbox{\boldmath $\delta$}\mbox{\boldmath $\vartheta$}.
\end{equation}
The \textsl{Hamiltonian vector field} is defined by
\begin{equation}
	\iota_{X_{f}}\bomega = -\mbox{\boldmath $\delta$}f,
\end{equation}
and the \textsl{BV bracket} we define by
\begin{equation}
	\left\{f,g\right\}_{BV} = X_{f}g.
\end{equation}
Then, direct computation gives the following local expression
of the BV bracket on the mapping space,
\begin{equation}
	\left\{f,g\right\}_{BV} = (-1)^{d-|q|}\int_{\calX}\left[\frac{f\overleftarrow{\delta}}{\delta \bbq^{i}}\mu\frac{\overrightarrow{\delta}g}{\delta \bbp_{i}} + (-1)^{d(1+|q|)}\frac{f\overleftarrow{\delta}}{\delta \bbp_{i}}\mu\frac{\overrightarrow{\delta}g}{\delta \bbq^{i}}\right].
\end{equation}
We can prove the following identities of the graded Poisson bracket of 
degree $1$,
\begin{eqnarray}
\mbv{f}{g}&=&-(-1)^{(|f|+1)(|g|+ 1)} \mbv{g}{f},
\\
\mbv{f}{g  h}&=&\mbv{f}{g} h + (-1)^{(|f|+1)|g|} g \mbv{f}{h},
\\
\mbv{f}{\mbv{g}{h}} &=& \mbv{\mbv{f}{g}}{h}
+(-1)^{(|f|+1)(|g|+1)}\mbv{g}{\mbv{f}{h}}.
\end{eqnarray}

\begin{remark}
We list up the degrees of the defined objects.

A vector field has
$|X| = -d + |X^{i}| + d - |\bbz^{i}| = |X^{i}| - |\bbz^{i}|$. 
An interior product has $|\iota_{X}| = -d + |X^{i}| + d - (|\bbz^{i}| + 1) 
= |X^{i}| - |z^{a}| - 1 = |X| - 1$, 
since $|\delta \bbz^{i}| = |\bbz^{i}| + 1$. 
A symplectic structure has 
$|\bomega| = -d + 1 + |\bbq^{i}| + 1 + |\bbp_{i}| = 1$. 
A differential on a function has $|\mbox{\boldmath $\delta$} f| 
= -d + 1 + |\bbz^{i}| + |f| + d - |\bbz^{i}| = 1 + |f|$. 
A Hamiltonian vector field has $|X_{f}| = 1 + |f|$ and 
therefore $|\iota_{X_{f}}| = |f|$.
\end{remark}

\subsection*{Acknowledgments}
\noindent
The authors would like to thank T.~Asakawa, B.~Jur\v{c}o, Y.~Kaneko, H.~Muraki, Y.~Maeda and U.~Carow-Watamura for helpful discussions and valuable comments.
M.A.~Heller is supported by Japanese Government (MONBUKAGAKUSHO) Scholarship
and
N.~Ikeda is supported by the research promotion program grant 
at Ritsumeikan University.

\newcommand{\bibit}{\sl}



\end{document}